\documentclass{article}

\usepackage{float}          % boxed figures
\usepackage{mathpartir}     % math paragraphs and inference rules
\usepackage{pstricks}
\usepackage{pst-node}
\usepackage{wrapfig}

\usepackage{polys-metavars}
\usepackage{polys-syntax}
\usepackage{polys-functions}
\usepackage{comparison}

\renewcommand{\land}{\mathrel{\&}}

\floatstyle{boxed}
\restylefloat{figure}

\title{Expressiveness of Generic Process Shape Types}
\date{
    { \normalsize Heriot-Watt  University } \\[2mm]
    March 31, 2010
    \footnote{We hereby grant the reader a perpetual, non-exclusive license
    to re-distribute this article.}
}

\author{Jan~Jakub\r{u}v \and J.~B.~Wells 
}
\newtheorem{theorem}{Theorem}[section]
\newtheorem{property}{Property}[section]
\newtheorem{example}{Example}[section]

\newenvironment{fakedexample}
    {   \refstepcounter{example} \vspace{2mm} \noindent
        \textit{Example \theexample.} }
    { \vspace{1mm} }

\begin{document}
%%fakesection{Title Page}
    
\maketitle
%\pagestyle{myheadings}
%\markboth{\hfill\today}{\today\hfill}
\vspace{-2ex}

\begin{abstract}
    Shape types are a general concept of process types which work for many
    process calculi.
    We extend the previously published \polyS system of shape types to
    support name restriction.
    We evaluate the expressiveness of the extended system by showing that
    shape types are more expressive than an implicitly typed $\pi$-calculus
    and an explicitly typed Mobile Ambients.
    We demonstrate that the extended system makes it easier to enjoy
    advantages of shape types which include polymorphism, principal typings,
    and a type inference implementation.
\end{abstract}

\vspace{-2ex}
\section{Introduction}

Many type systems for many process calculi have been developed to statically
guarantee various important properties of processes.
Types differ among these systems and their properties, such as
soundness, have to be proved separately for each system.
\emph{Shape types} are a general concept of polymorphic process types which
can express and verify various properties of processes.
\polyS \cite{Mak+Wel:ESOP-2005,Mak+Wel:PolyStar-2004} is a general framework
which, for a wide range of process calculi, can be instantiated to make
ready-to-use sound type systems which use shape types.
Only rewriting rules satisfying common syntactic conditions are
needed for instantiating \polyS.

Many process calculi share semantically equivalent constructions, such as,
\emph{parallel composition} (``$\oI$''), prefixing a process with an action
(sometimes called a \emph{capability}) (``$.$''), and \emph{name
restriction} (``$\nu$'').
Specific calculi differ mainly in the syntax and semantics of actions
(capabilities).
\metaS \cite{Mak+Wel:ESOP-2005,Mak+Wel:PolyStar-2004} is metacalculus which
fixes semantics of the shared constructions and provides a way to describe
syntax and semantics of actions by a description $\xRS$ of rewriting rules.
Given $\xRS$, \metaS makes the calculus $C_\xRS$ and \polyS makes the type
system $S_\xRS$ for $C_\xRS$.
$\xRS$ can describe many calculi including, e.g., the $\pi$-calculus,
Mobile Ambients, numerous variations of these, and other systems.
All instantiations of \polyS share \emph{shape predicates} which describe
allowed syntactic configurations of \metaS processes.
\emph{Shape ($\xRS$-)types} of $S_\xRS$ are shape predicates whose
meaning is guaranteed by a simple test to be closed under rewriting with
$\xRS$.
Every $S_\xRS$ has desirable properties such as subject reduction, the
existence of principal typings \cite{Wells:ICALP-2002}, and an already
implemented type inference algorithm%
\footnote{\texttt{http://www.macs.hw.ac.uk/ultra/polystar} (includes a
web demonstration)}.

\subsection{Contributions}\label{sec:contributions}

This paper extends the \polyS system to support name restriction and also
proves \polyS shape types are more expressive than some previous systems for
specific calculi.
The contributions are as follows.
(1) Sec.~\ref{sec:star} presents the extended \polyS system.
Sections ~\ref{sec:pi}, \ref{sec:ma} 
    show (2) how to easily use shape types with well-known calculi (%
    the $\pi$-calculus \cite{Mil+Par+Wal:IC-1992,Mil:CMS-1999}, 
    Mobile Ambients \cite{Car+Gor:FoSSaCS-1998}),
(3) demonstrate polymorphic abilities of shape types, and
(4) prove that shape types are more expressive than predicates of two
type systems (%
    \emph{implicitly} typed $\pi$-calculus \cite{Tur:PhD-1995},
    \emph{explicitly} typed Mobile Ambients \cite{Car+Gor:POPL-1999})
custom designed for the above calculi.
Finally, (5) we advocate a generic notion of shape types and show that
they can be used instead of predicates of many other systems.
We consider contributions (4) \& (5) to be the main contribution of the
paper.

Contribution (2) shows how to use \polyS and shape types without needing to
fully understand all the details of the underlying formalism.
Thus it helps to bridge over the problem of complexity of \polyS
which is inevitably implied by its high generality and which has been
daunting to some readers of earlier papers.
Contribution (3) shows an aspect of shape types which is not common for other
systems.
An accompanying technical report \cite{Jak+Wel:ShapeTypes-2009} (TR),
which extends this paper and contains proofs of main theorems,
additionally shows how to use shape types for \emph{flow analysis} of
BioAmbients and proves its superior expressiveness to an earlier flow
analysis system \cite{Nie+Nie+Pri+Ros:ENTCS-2007-v180n3}.
This work was left out for space reasons.
For all the three systems we have proven not only that shape types are more
expressive but also that they can be used to achieve exactly the same
results as the original systems which might be important for some of
their applications.
We believe that the diversity of the mentioned systems and their intended
applications provides a reasonable justification for contribution (5).

\subsection{Notations and Preliminaries}\label{sec:notation}

Let $i$, $j$, $k$ range over natural numbers.
$\powerfin{U}$ is the set of all finite subsets of a set $U$,
``$\backslash$'' denotes set subtraction. 
Let $u\mapsto v$ be an alternate pair notation used in functions.
$\extend{f}{u\mapsto v}$ stands for the function that maps $u$ to $v$ and
other values as $f$.  Moreover, $\map{U}{V}$ ($\mapfin{U}{V}$) is the set of
all (all finite) functions $f$ with 
$\dom{f}\subseteq U$ and $\rng{f}\subseteq V$.

\section{Metacalculus \metaS and Generic Type System \polyS}
\label{sec:star}

%\metaS~\cite{Mak+Wel:ESOP-2005,Mak+Wel:PolyStar-2004} has changed: names are
%now built from basic names (Sec.~\ref{sec:star/calculus}) which makes a
%simple \polyS name restriction rule (Sec.~\ref{sec:poly/types}) do the right
%thing.

\subsection{General Syntax of Processes}\label{sec:star/calculus}

\newcommand{\FIGmetasyntax}{
\begin{figure}[t]
\vspace{-1mm}
\small
    \begin{pstgrammar}
        \grmclass{\xa,\xab}{BasicName}{ 
            \name{a} \grmOr \name{b} \grmOr \cdots
             \grmOr \name{in} \grmOr \name{out} \grmOr \name{open} \grmOr \cdots
             \grmOr \amb{}{} \grmOr \sBullet \grmOr \cdots
        } \\
        \grmclass{\x,\xy}{Name}{\xa^\xn} \\
        \grmclass{\xF}{Form}{ \x_0\ldots\x_k } \\
        \grmclass{\xM}{Message}{ \xF \grmOr \sNull \grmOr \xM_0.\xM_1 } \\
        \grmclass{\xE}{Element}{ \x \grmOr \sIn{\x_1,\ldots,\x_k}
            \grmOr \sOut{\xM_1,\ldots,\xM_k} } \\
        \grmclass{\xA}{Action}{ \xE_0\ldots\xE_k } \\
        \grmclass{\xP,\xQ}{Process}{ 
            \pNull \grmOr
            \xA.\xP \grmOr 
            (\xP \oI \xQ) \grmOr
            \pNu{\x}{\xP} \grmOr 
            \pBang{\xP} 
        }
    \end{pstgrammar}
\normalsize
\vspace{-2mm}
\caption{Syntax of \metaS processes.}
\label{fig:meta/syntax}
\end{figure}
}

\newcommand{\FIGmetastreq}{
\begin{figure}[t]
\small
\begin{displaymath}
\begin{array}{lll}
    \infAxiom{ \streq{\pPar{\xP}{\xQ}}
                     {\pPar{\xQ}{\xP}} \qquad }
&
    \infAxiom{ \streq{\pPar{\xP}{(\pPar{\xQ}{\xR})}}
                     {\pPar{(\pPar{\xP}{\xQ})}{\xR}} \qquad }
&
    \infAxiom{ \streq{\pPar{\xP}{\pNull}}{\xP} }
\\
    \infAxiom{ \streq{\pNull}{\pBang\pNull} }
&
    \infAxiom{ \streq{\pNu{\xx}{\pNu{\xy}{\xP}}}
                     {\pNu{\xy}{\pNu{\xx}{\xP}}} }
&
    \infAxiom { \streq{\oB\xP}{\xP\oI\oB\xP} }
\end{array}
\end{displaymath}
\vspace{-2mm}
\begin{displaymath}
    \infCondAxiom
          { \x\not\in\fn{\xA}\cup\bn{\xA} }
          { \streq{\xA.\pNu{\x}{\xP}}{\pNu{\x}{\xA.\xP}} }
\qquad
    \infCondAxiom
          { \x\not\in\fn{\xP} }
          { \streq{\pPar{\xP}{\pNu{\x}{\xQ}}}
                {\pNu{\x}{(\pPar{\xP}{\xQ})}} }
\end{displaymath}
\vspace{-3mm}
\normalsize
\caption{\metaS structural equivalence (structural rules omitted).}
\label{fig:meta/streq}
\end{figure}
}

\FIGmetasyntax

\metaS process syntax, presented in Fig.~\ref{fig:meta/syntax}, allows
embeddings of many calculi.
A name $\xa^i$ is a pair of a \emph{basic name} $\xa$ and a natural
number $i$.
The basic part of a name $\xx$ is denoted $\base{\xx}$, that is,
$\base{\xa^i}=\xa$.
When $\alpha$-converting, we preserve the basic name and change the
number.
We write $\xa$ instead of $\xa^0$ when no confusion can arise.

Processes are built from the null process ``$\pNull$'' by prefixing with an
action (``.''), by parallel composition (``$\oI$''), by name restriction (``$\nu$''), and by
replication (``$\oB$'').
Actions can encode prefixes from various calculi such as
$\pi$-calculus communication actions, Mobile Ambients
capabilities, or ambient boundaries.
The abbreviation ``$\amb{\x_1\ldots\x_k}{\xP}$'', which further supports
ambient syntax, stands for ``$\x_1\ldots\x_k\sAmb.\xP$''
($\amb{}{}$ is a single name).

\renewcommand{\infAxiom}[1]{\inferrule*{}{#1}}
\renewcommand{\infCondAxiom}[2]{#2\mbox{  if  }#1}
\FIGmetastreq

Process constructors have standard semantics.
``$\pNull$'' is an inactive process, ``$\xA.\xP$''
executes the action $\xA$ and continues as $\xP$, ``$\xP\oI\xQ$''
runs $\xP$ and $\xQ$ in parallel,
``$\pNu{\xx}{\xP}$'' behaves as $\xP$ with private name $\xx$ (i.e., $\xx$
differs from all names outside $\xP$), and ``$\oB\xP$''
acts as infinitely many copies of $\xP$ in parallel (``$\xP\oI\xP\oI\cdots$'').
Let ``$.$'' and ``$\nu$'' bind more tightly than ``$\oI$''.
These constructors have standard properties given by structural equivalence
$\SYMstreq$ (Fig.~\ref{fig:meta/streq}), e.g., ``$\oI$'' is
commutative, adjacent ``$\nu$'' can be interchanged, etc.
In contrast, the semantics of actions is defined by
instantiating \metaS (see below).
Currently, \metaS does not support the choice operator ``$+$'' as a built in
primitive.
However, ``$\xP + \xQ$'' can be encoded as ``$\n{ch}.(\xP\oI\xQ)$'' provided
rewriting rules are extended to use this encoding.

All occurrences $\x$ in ``$\pNu{\x}{\xP}$'' are
\mbox{($\nu$-)bound}.
When the action $\xA$ contains an element ``$\sIn{\x_1,\ldots,\x_\xnk}$'' then
all occurrences of the $\x_\xn$'s in ``$\xA.\xP$'' as well as in $\xA$ on its
own are called \mbox{(input-)bound}.
An occurrence of $\x$ that is not bound is free.
The occurrence of $\xa$ in $\xa^i$ is bound (resp. free) when this
occurrence of $\xa^i$  is.
A bound occurrence of $\xa^i$ can be $\alpha$-converted 
only to $\xa^j$ with $\xa$ the same.
We identify $\alpha$-convertible processes.
The set of free names of $\xP$ is denoted $\fn{\xP}$.
The set $\fbn{\xP}$ (resp. $\ibn{\xP}$, $\nbn{\xP}$) contains free (resp.
input-bound, $\nu$-bound) basic names of $\xP$.
The set of bound names of $\xA$ is written $\bn{\xA}$.

A process $\xP$ is \defthis{well scoped} when \rulename{(W1)} 
$\fbn{\xP}$, $\ibn{\xP}$, and $\nbn{\xP}$ do not overlap,
\rulename{(W2)} nested input binders do not bind the same basic name, and
\rulename{(W3)} no action contains an input-binding of a basic name more
than once.
These conditions are important for type inference.
We allow only well scoped processes.

A \metaS substitution $\xSub$ is a finite function from $\grmset{Name}$
to $\grmset{Message}$.
Application of $\xSub$ to $\xP$, written $\sub{\xP}$, behaves as
usual except the following. %two situations.
(1) It places a special name ``$\sBullet$'' at positions that would
otherwise be syntax errors (e.g., 
$\sub[\{ \n{x}\mapsto \cab{out}{\n{b}} \}]
     {(\cab{in}{\n{x}}.\pNull)} = \cab{in}{\sBullet}.\pNull$).
(2) When a composed message $\xM$ is substituted for a single name
action $\n{x}$ in ``$\n{x}.\xP$'', then $\xM$'s components are pushed from
right to left onto $\sub{\xP}$ (e.g.,
$\sub[\{ \n{x}\mapsto(\n{a}.\n{b}).\n{c} \}]
     {(\n{x}.\pNull)} = \n{a}.\n{b}.\n{c}.\pNull$).
%For case (2), an auxiliary operation $\splice{\xM}{\xP}$ is defined
%in Fig.~\ref{fig:meta/streq}.
The full definition of $\sub{\xP}$ is in the TR.

%The structural equivalence relation $\SYMstreq$ is the smallest
%equivalence that satisfies the rules in
%Fig.~\ref{fig:meta/streq} and is congruent with the \metaS process
%constructors.

\subsection{Instantiations of \metaS}\label{sec:star/instantiate}

\newcommand{\FIGmetarwrules}{
\begin{figure}[t]
\small
\labeledHeader{Syntax of \metaS rewriting rules:}
\\[-7mm]
\begin{minipage}[t]{6.5cm}
\begin{pstgrammar}
    \grmclass{\xNVx,\xNVy}{NameVar}{ 
        \var{a} \grmOr 
        \var{b} \grmOr 
        \var{c} \grmOr 
        \cdots } 
\\
    \grmclass{\xMV}{MessageVar}{ 
        \var{M} \grmOr 
        \var{N} \grmOr 
        \cdots } 
\\
    \grmclass{\xETpl}{ElementTempl}{ 
        \x \grmOr 
        \xNV \grmOr 
        \sIn{\xNV_1,\ldots,\xNV_k} \grmOr 
        \sOut{\xMV_1,\ldots,\xMV_k} } 
\\
    \grmclass{\xATpl}{ActionTempl}{ 
        \xETpl_0\;\xETpl_1 
        \ldots \xETpl_k } 
\\
    \grmclass{\xPTplP,\xPTplQ}{ProcessTempl}{ 
        \xPV \grmOr 
        \xATpl.\xPTpl \grmOr 
        \pNull \grmOr 
        (\pPar{\xPTplP}{\xPTplQ}) \grmOr
        \sTemplSubst\,\xPV }
\\
    \grmclass{\xRule}{Rule}{ 
        \sReduce{\xPTplP}{\xPTplQ} \grmOr 
        \sActive{\xPV}{\xPTpl} 
    } 
\\
    \grmeqclass{\xRS}{RuleSet}{ \powerfin{\mathsf{Rule}} }
\end{pstgrammar}
\end{minipage}
\begin{minipage}[t]{5cm}
\begin{pstgrammar}
    \grmclass{\xPV}{ProcessVar}{ 
        \var{P} \grmOr 
        \var{Q} \grmOr 
        \var{R} \grmOr 
        \cdots }
\\
    \grmclass{\xSubstute}{Substitute}{ 
        \xNV \grmOr 
        \xMV } 
\end{pstgrammar}
\end{minipage}
\vspace{-3mm}
\labeledHline{Semantics of \metaS rewriting rules:}
\renewcommand{\arraystretch}{1.2}
\begin{displaymath}
\begin{array}{ll}
    \infRule
        { \sReduce{\xPTplP}{\xPTplQ} \in \xRS }
        { \reduce{\TermI{\xPTplP}}{\TermI{\xPTplQ}} }
  &
    \infRule
        { \reduce{\xP}{\xQ} }
        { \reduce{\pNu{\x}{\xP}}{\pNu{\x}{\xQ}} }
\\
    \infRule
        {   \streq{\xP'}{\xP} \land
            \reduce{\xP}{\xQ} \land 
            \streq{\xQ}{\xQ'} }
        {   \reduce{\xP'}{\xQ'} }
    \quad
  &
    \infRule
        { \reduce{\xP}{\xQ} }
        { \reduce{\xP\oI\xR}{\xQ\oI\xR} }

\\
\multicolumn{2}{c}{
    \infRule
        {   \sActive{\xPV}{\xPTpl} \in \xRS \land
            \reduce{\xP}{\xQ} }
        {   \reduce{ \TermISpec{\xPTpl}{\xPV\mapsto\xP} }
                   { \TermISpec{\xPTpl}{\xPV\mapsto\xQ} } }
}
\end{array}
\end{displaymath}
\vspace{-1mm}
\normalsize
\caption{Syntax and semantics of \metaS reduction rule descriptions.}
\label{fig:meta/rwrules}
\end{figure}
}

\renewcommand{\infAxiom}[1]{\inferrule*{}{#1}}
\renewcommand{\infRule}[2]{#1 \,\Rightarrow\, #2}

\metaS provides syntax to describe rewriting rules that give meaning to
actions and also defines how these rules yield a rewriting relation on
processes.
The syntax is best explained by an example.
The following rule description
    (in which ``$\oLeft\var{x}\oAssign\var{n}\oRight\var{Q}$''
    describes substitution application)
%
%\vspace{-1mm}
\begin{displaymath}
\sReduce{
    \sChOut{\var{c}}{\var{n}}.\var{P} \oI
    \sChIn{\var{c}}{\var{x}}.\var{Q} 
}{
    \var{P}\oI\oLeft
        \var{x}\oAssign\var{n}
    \oRight\var{Q}
}
\end{displaymath}
%\vspace{-1mm}
%
%\noindent
directly corresponds to the standard $\pi$-calculus communication
rule
``$
    \sChOut{c}{n}.P \oI
    \sChIn{c}{x}.Q 
    \Rightarrow
    P\oI\sub[\{x\mapsto n\}]{Q}
$''.
The circle-topped letters stand at the place of name, message, and process
metavariables.
Given a set $\xRS$ of rule descriptions in the above syntax, \metaS
automatically infers the rewriting relation $\SYMreduce$ which
incorporates structural equivalence and congruence rules (e.g.,
``$\reduce{P}{Q}\Rightarrow\reduce{\pNu{x}{P}}{\pNu{x}{Q}}$'').
A rules description instantiates \metaS to a particular calculus, e.g.,
the set $\xRS$ containing only the above rule description
instantiates \metaS to the $\pi$-calculus.

Further examples of \metaS instantiations are given in
Sec.~\ref{sec:pi/instatiate} and \ref{sec:ma/instatiate}.
A rule description can also contain a concrete \metaS name (e.g.
``$\n{out}$'') when an exact match is required.
We require that these names are never bound in any process.
Complete definitions of the syntax of rewriting rules and of the
rewriting relation $\SYMreduce$ is left to the TR
\cite[Sec.~2.2]{Jak+Wel:ShapeTypes-2009}.

\subsection{\polyS Shape Predicates and Types for \metaS}\label{sec:poly/types}

A \emph{shape predicate} describes possible structures of process
syntax trees.
When a rewriting rule from $\xRS$ is applied to a process, its syntax tree
changes, and sometimes the new syntax tree no longer satisfies the same
shape predicates.
All \polyS ($\xRS$-)types are shape predicates that describe process
sets closed under rewriting using $\xRS$.
For feasibility, types are defined via a syntactic test that enforces
rewriting-closedness.
Intuitively, the syntactic test tries to apply the rules from $\xRS$ to all
active positions in a shape graph and checks whether all the edges newly
generated by this application are already present in the graph.
Further restrictions are used to ensure the existence of principal
typings.

\newcommand{\FIGpolyshapesyntax}{
\begin{figure}[t]
\small
\labeledHeader{Syntax of \polyS shape predicates:}
\\[-7mm]
\begin{minipage}[t]{6.5cm}
    \begin{pstgrammar}
        \grmclass{\xFT}{FormType}{ \xa_0\ldots\xa_k } \\
        \grmeqclass{\xFS}{Form\mkern-3muTypeSet}{
            \powerfin{\grmset{FormType}}            
        } \\
        \grmclass{\xMT}{MessageType}{ 
            \sStar{\xFS} \grmOr
            \xa
        } \\
        \grmclass{\xET}{ElementType}{ \xa
            \grmOr \sIn{\xa_1,\ldots,\xa_k} \grmOr
        } \\
        \grmcont{ \sOut{\xMT_1,\ldots,\xMT_k} } 
    \end{pstgrammar}
\end{minipage}
\begin{minipage}[t]{4.5cm}
    \begin{pstgrammar}
        \grmclass{\xAT}{ActionType}{ \xET_0\;\xET_1\ldots \xET_k } \\
        \grmclass{\xNode}{Node}{ 
            \node{X} \grmOr 
            \node{Y} \grmOr 
            \node{Z} \grmOr 
            \cdots 
        } \\
        \grmclass{\xEdge}{Edge}{ \xNode_0\oEdge{\xAT}\xNode_1 } \\
        \grmeqclass{\xG}{ShapeGraph}{ \powerfin{\grmset{Edge}} } \\
        \grmclass{\xS}{ShapePredicate}{ \sShape{\xG}{\xNode} }
    \end{pstgrammar}
\end{minipage}
\vspace{-3mm}

\labeledHline{Rules for matching \metaS entities against shape predicates:}
\begin{displaymath}
\setlength{\arraycolsep}{2mm}
\begin{array}{lll}
    \infAxiom 
          { \shapes{\xa^\xn}{\xa} }
  &
    \infAxiom
          { \shapes{\sIn{\xa_1^{i_1},\ldots,\xa_k^{i_k}}}
                   {\sIn{\xa_1,\ldots,\xa_k}} }
  &
    \infRule
          { (\shapes{\xM_0}{\xFS}
             \land\, \shapes{\xM_1}{\xFS}) }
          { \shapes{\xM_0.\xM_1}{\xFS} }
\\
    \infAxiom
          { \shapes{\pNull}{\xFS} }
  &
    \infRule
          { (\shapes{\xF}{\xFT} \land \xFT\in\xFS) }
          { \shapes{\xF}{\xFS} }
  &
    \infRule
               %{ \xM\ne\xa \land
               { (\xM\not\in\grmset{Name} \land\, 
                 \shapes{\xM}{\xFS}) }
               { \shapes{\xM}{\sStar{\xFS}} }
%\\
%    \infRule
%          { (\forall i\!. 0<i\le k\!:\ 
%            \shapes{\xM_i}{\xMT_i}) 
%          }
%          { \shapes{\sOut{\xM_1,\ldots,M_k}}{\sOut{\xMT_1,\ldots,\xMT_k}} }
%\\
%     \infRule
%           { (\forall i\le k\!:\ 
%             \shapes{\xE_i}{\xET_i})
%           }
%           { \shapes{\xE_0 \ldots \xE_k}{\xET_0 \ldots \xET_k} }
\end{array}
\end{displaymath}
\begin{displaymath}
\setlength{\arraycolsep}{2mm}
\begin{array}{ll}
     \infRule
           { (\forall i\le k\!:\ 
             \shapes{\xE_i}{\xET_i})
           }
           { \shapes{\xE_0 \ldots \xE_k}{\xET_0 \ldots \xET_k} }
  &
\\
    \infRule
          { (\forall i\!: 0<i\le k \land\, \shapes{\xM_i}{\xMT_i}) }
          { \shapes{\sOut{\xM_1,\ldots,M_k}}{\sOut{\xMT_1,\ldots,\xMT_k}} }
\end{array}
\end{displaymath}
\begin{displaymath}
\setlength{\arraycolsep}{1mm}
\begin{array}{lr}
    \begin{array}{l}
        \infAxiom{ \shapes{\pNull}{\xS} }
    \\
        \infRule
              { \shapes{\xP}{\xS} }
              { \shapes{\pNu{\xx}{\xP}}{\xS} } 
    \\
        \infRule
              { \shapes{\xP}{\xS} }
              { \shapes{\oB\xP}{\xS} } 
    \end{array}
  &
    \begin{array}{l}
        \infRule
              { (\shapes{\xP}{\xS}
                 \land\, \shapes{\xQ}{\xS}) }
              { \shapes{\xP\oI\xQ}{\xS} }
    \\
        \infRule{ ((\xNode_0\oEdge{\xAT}\xNode_1) \in \xG \land\,
            \shapes{\xA}{\xAT}  \land\,
            \shapes{\xP}{\sShape{\xG}{\xNode_1}})}
        { \shapes{\xA.\xP}{\sShape{\xG}{\xNode_0}} }
    \end{array}
\end{array}
\end{displaymath}
\vspace{-2mm}
\normalsize
\caption{Syntax and semantics of \polyS shape predicates.}
\label{fig:poly:shape/syntax}
\end{figure}
}

\renewcommand{\infAxiom}[1]{#1}
\renewcommand{\infRule}[2]{#1\mathop{\Rightarrow}#2}
\FIGpolyshapesyntax

Fig.~\ref{fig:poly:shape/syntax} defines shape predicate syntax.
Action types are similar to actions except that action types are built from
basic names instead of names, and compound messages are described
up to commutativity, associativity, and repetitions of their parts.
Thus an action type describes a set of actions. 
A shape predicate $\sShape{\xG}{\xNode}$ is a directed finite graph 
with root $\xNode$ and with edges labeled by action types.
A process $\xP$ matches $\xS$ when $\xP$'s syntax tree is a ``subgraph''
of $\xS$.
Shape predicate can have loops and thus describe syntax trees of
arbitrary height.

Fig.~\ref{fig:poly:shape/syntax} also describes
matching \metaS entities against shape predicates.
The rule matching actions against action types also matches
forms against form types.
Matching entities against types does not depend on $\xRS$,
i.e., it works the same in any \metaS instantiation.
The \defthis{meaning} $\means{\xS}$ of the shape predicate $\xS$ is the
set $\{ \xP|\shapes{\xP}{\xS} \}$ of all processes matching $\xS$.

A shape predicate $\xS$ is \defthis{semantically closed} w.r.t.\
a rule set $\xRS$ when $\means{\xS}$ is closed under
$\xRS$-rewritings, i.e.,
when $\shapes{\xP}{\xS}$ and $\reduce{\xP}{\xQ}$ imply
$\shapes{\xQ}{\xS}$ for any $\xP$ and $\xQ$.
Because deciding semantic closure w.r.t.\ an
arbitrary $\xRS$ is nontrivial, we use an easier-to-decide property,
namely \defthis{syntactic closure}, which by design is
algorithmically verifiable.
\emph{$\xRS$-types} are shape predicates syntactically closed w.r.t.\
$\xRS$.
A type $\xS$ of $\xP$ is a \emph{principal typing} of $\xP$ when
$\means{\xS}\subseteq\means{\xS_0}$ for any other type $\xS_0$ of $\xP$.
There are \emph{width} and \emph{depth} restrictions to ensure principal
typings.
Details are left to
    our TR \cite[Sec.~2.4]{Jak+Wel:ShapeTypes-2009}.

\subsection{Proving Greater Expressiveness of \polyS}
\label{sec:general}
\label{sec:general/terminology}
\label{sec:general/calculus}
\label{sec:general/system}

We now discuss how to consider some process calculus $C$ and its type
system $S_C$ and prove the greater expressiveness of the related
\metaS and \polyS instantiations.
Sections \ref{sec:pi} and \ref{sec:ma} follow this approach.
Usually $S_C$ defines predicates (ranged over by $\gxT$) which represent
properties of processes (ranged over by $\gxP$) of $C$.
Then $S_C$ defines the relation $\gType{\gxP}{\gxT}$ which represents
statements ``$\gxP$ has the property $\gxT$'' and which is preserved under
rewriting of $\gxP$ in $C$.
The \metaS description $\xRS$ of $C$'s rewriting rules gives us the calculus
$C_\xRS$ and its shape type system $S_\xRS$.

Firstly we need to set up a correspondence between $C$ and $C_\xRS$, that
is, we need an encoding $\encode{\cdot}$ of processes $\gxP$ into \metaS
which preserves $C$'s rewriting relation $\SYMgRew$.
The following property, which is usually
easy to prove, formulates this modulo $\SYMgStrEq$ because structural
equivalences of different calculi might differ.
\begin{property}\label{thm:general/encoding+correct}
When $\gRew{\gxPP}{\gxPQ}$ then $\exists\gxPP',\gxPQ'$ such
that 
\begin{math}
    \gStrEq{\gxPP}{\gxPP'} \land
    \reduce{\encode{\gxPP'}}{\encode{\gxPQ'}} \land
    \gStrEq{\gxPQ'}{\gxPQ}.
\end{math}
When $\reduce{\encode{\gxPP}}{\xP_1}$ then $\exists\gxPQ$ such that
\begin{math}
    \gRew{\gxPP}{\gxPQ} \land
    \gStrEq{\encode{\gxPQ}}{\xP_1}.
\end{math}
\end{property}

Predicates $\gxT$ of $S_C$ are commonly preserved under renaming of
bound basic names, that is,  $\gType{(\nu\xx)\gxP}{\gxT}$ usually implies
$\gType{(\nu\xa^0)(\sub[\{\xx\mapsto\xa^0\}]{\gxP})}{\gxT}$ (for $\xa$
not in $\gxP$).
Predicates of similar systems can not be directly translated to \polyS shape
types with the corresponding meaning because shape types do not have this
property.
In other words, the difference in handling of bound names between \polyS and
other systems makes some straightforward embeddings impossible.

We investigate two reasonable ways to embed $S_C$ in $S_\xRS$, that
is, to decide $\gType{\gxP}{\gxT}$ using $S_\xRS$'s relation
``$\vdash$''.
(1) In Sec.~\ref{sec:ma/embedding} about Mobile Ambients, we translate
$\gxT$ together with information about bound basic names of $\gxP$ into a
shape type.
(2) In Sec.~\ref{sec:pi/embedding} about the $\pi$-calculus, we show how to
decide $\gType{\gxP}{\gxT}$ by a simple check on a principal shape
type of $\gxP$.
The fact that both embeddings of predicates $\gxT$ depend on a process
$\gxP$ is not a limitation because $\gxP$ is known for desirable
applications like type checking.

We stress that these embeddings serve the theoretical purpose of proving
greater expressiveness and are not necessary for a practical use of shape
types.
When $S_C$ is designed to verify a certain fixed property of processes which
can be expressed as a property of shape types, then we can use $S_\xRS$
directly for the same purposes as $S_C$ without any embedding.
We show how to do this for the two systems in
Sec.~\ref{sec:pi/instatiate} and \ref{sec:ma/instatiate}.
We can also design a property of processes directly on shape types without
any reference to another analysis system.
Our TR \cite[Sec. 3]{Jak+Wel:ShapeTypes-2009} discusses this further.

\subsection{Discussion}
\label{sec:star/discussion}

\polyS presented above extends the previously published
\polyS \cite{Mak+Wel:ESOP-2005} with name restriction.
The previously published system \cite{Mak+Wel:ESOP-2005} supports
restriction only in \metaS but no processes with $\nu$ are typable in \polyS
instantiations.
An earlier attempt in a technical report \cite{Mak+Wel:PolyStar-2004} to
handle name restriction was found inconsistent
\cite[Sec.~3.2-4]{Jak:SYR-2009} and furthermore inadequate
\cite[Sec.~4]{Jak:SYR-2009} to carry out the proofs of greater
expressiveness in sections \ref{sec:pi} and \ref{sec:ma}.

The difficulty with name restriction is because a shape
type represents a syntactic structure of a process, and thus presence of
bound names in a process has to be somehow reflected by a shape graph.
Because bound names can be $\alpha$-renamed, \polyS needs to establish a
connection between positions in a process and a shape graph which is
preserved by $\alpha$-conversion.
This connection is provided by basic names which are the key concept of
name restriction handling in this paper.
For example, for the action ``$\n{a}\sOut{\n{a}}$'' there is the
corresponding action type ``$\n{a}\sOut{\n{a}}$'' in its shape type.
When the name $\n{a}$ were $\nu$-bound and $\alpha$-renamed to some other
name then the correspondence between the action in the process and the
action type would be lost.
This problem is solved by building shape types from basic names which are
preserved under $\alpha$-conversion.

The handling of input-bound names in the previous \polyS was reached by
disabling their $\alpha$-conversion which is possible under certain
circumstances.
But $\alpha$-conversion of $\nu$-bound names can not be avoided and thus a
new approach has been developed.

\section{Shape Types for the $\pi$-calculus}
\label{sec:pi}

\newcommand{\PARinstpi}{
\small
\begin{displaymath}
\begin{array}{l}
    \mathcal{P} = \bigcup_{k=0}^{\infty} \big\{ 
    \oReduce\oLeft\,
        \sChOut{\var{c}}{\var{M}_1,\ldots,\var{M}_k}.\var{P} \!\oI\!
        \sChIn{\var{c}}{\var{a}_1,\ldots,\var{a}_k}.\var{Q} 
    \ \oArr
        \var{P}\!\oI\!\oLeft
            \var{a}_1\!\oAssign\var{M}_1,\ldots,
            \var{a}_k\!\oAssign\var{M}_k
        \oRight\var{Q}
        \,\oRight
    \big\}
\end{array}
\end{displaymath}
\normalsize
}

\subsection{A Polyadic $\pi$-calculus}
\label{sec:pi/calculus}

\newcommand{\FIGtpisyntaxsemantics}{
\begin{figure}[t]
\small
\labeledHeader{Syntax of the $\pi$-calculus processes:}
\begin{pstgrammar}
    \grmeqclass{\cxc,\cxx,\cxy}{PiName}{
        \grmset{Name}\setminus{\{\sBullet\}}}     \\
    \grmclass{\cxA}{PiAction}{
        \cIn{\cxc}{\cxx_1,\ldots,\cxx_k} \grmOr
        \cOut{\cxc}{\cxx_1,\ldots,\cxx_k} 
    } \\
    \grmclass{\cxP}{PiProcess}{ 
        \pNull \grmOr
        (\cxPP \oI \cxPQ) \grmOr
        \cxA.\cxP \grmOr
        \oB \cxP \grmOr
        \gNu{\cxx}{\cxP}
    } 
\end{pstgrammar}
\setlength{\arraycolsep}{1mm}
\labeledHline{Rewriting relation of the $\pi$-calculus 
    ($\SYMbStreq$ is standard defined in TR
    \cite[Fig.~8]{Jak+Wel:ShapeTypes-2009}):}
\begin{displaymath}
\begin{array}{ll}
    \multicolumn{2}{l}{
        \cIn{\cxc}{\cxx_1,\ldots,\cxx_k}.\cxPP    \oI
        \cOut{\cxc}{\cxy_1,\ldots,\cxy_k}.\cxPQ \SYMtRewrite 
        \tSub[\{\cxx_1\mapsto\cxy_1,\ldots,\cxx_k\mapsto\cxy_k\}]{\cxPP}
        \oI \cxPQ
    }
\\[1mm]
    \infRule{ \tRewrite{\cxPP}{\cxPQ} }
            { \tRewrite{\cNu{\cxx}{\cxPP}}
                       {\cNu{\cxx}{\cxPQ}} }
&
        \tStreq{\cxPP'}{\cxPP} \land
        \tRewrite{\cxPP}{\cxPQ} \land
        \tStreq{\cxPQ}{\cxPQ'} 
        \,\Rightarrow\,
        \tRewrite{\cxPP'}{\cxPQ'} 
\\
    \infRule{ \tRewrite{\cxPP}{\cxPQ} }
            { \tRewrite{\cxPP\oI\cxPR}{\cxPQ\oI\cxPR} }
&   
\end{array}
\end{displaymath}
\vspace{-1mm}

\normalsize
\caption{The syntax and semantics of the $\pi$-calculus.}
\label{fig:pi/syntax+semantics}
\end{figure}
}

\newcommand{\FIGtpitypesrules}{
\begin{figure}[t]
\small
\labeledHeader{Syntax of \TPi types:}
\begin{pstgrammar}
    \grmclass{\cxTV}{PiTypeVariable}{ 
        \textsf{\i} \grmOr \textsf{\i'} \grmOr \textsf{\i''} \grmOr \cdots } \\
    \grmclass{\cxT}{PiType}{ \cxTV \grmOr 
        \cChT{\cxT_1,\ldots,\cxT_k} } \\
    \grmeqclass{\gxE}{PiContext}{ 
        \mapfin{\grmset{BasicName}}{\grmset{PiType}} } \\
\end{pstgrammar}
\labeledHline{Typing rules of \TPi:}
\begin{displaymath}
\begin{array}{llllllll}
    \infAxiom{\cType{\pNull}}
&   \infRule{ \cType{\cxPP} \land \cType{\cxPQ} }
            { \cType{\cxPP\oI\cxPQ} }
\\
    \infRule{ \cType{\cxP} }
            { \cType{\oB\cxP} } \quad
&   \infRule{ \cType[\extend{\cxE}{\base{\cxx}\mapsto\cxT}]{\cxP} }
            { \cType{\cNu{\cxx}{\cxP}} }
\end{array}
\end{displaymath}
\vspace{-2mm}
\begin{displaymath}
\begin{array}{llll}
    \infRule{ \cxE(\base{\cxc}) = \cChT{\cxT_1,\ldots,\cxT_k} \land
              \cType[\extend{\cxE}{\base{\cxx_1}\mapsto\cxT_1,\ldots,
                                   \base{\cxx_k}\mapsto\cxT_k}]{\cxP}
            }
            { \cType{\cIn{\cxc}{\cxx_1,\ldots,\cxx_k}.\cxP} }
\\
    \infRule{ \cxE(\base{\cxc}) = \cChT{\cxE(\base{\cxx_1}),\ldots,\cxE(\base{\cxx_k})}
              \land \cType{\cxP} }
            { \cType{\cOut{\cxc}{\cxx_1,\ldots,\cxx_k}.\cxP} }
\end{array}
\end{displaymath}
\vspace{-2mm}
\normalsize
\caption{Syntax of \TPi types and typing rules.}
\label{fig:pi/types+rules}
\end{figure}
}

The $\pi$-calculus \cite{Mil+Par+Wal:IC-1992,Mil:CMS-1999} is a process
calculus involving process mobility developed by Milner, Parrow, and Walker.
Mobility is abstracted as channel-based communication whose objects are
atomic names.
Channel labels are not distinguished from names and can be passed by
communication.
This ability, referred as \emph{link passing}, is the $\pi$-calculus
feature that most distinguishes it from its predecessors.
We use a polyadic version of the $\pi$-calculus which supports
communication of tuples of names.

\renewcommand{\infAxiom}[1]{#1}
\renewcommand{\infCondAxiom}[2]{#2 \mbox{\ \ if } #1}
\renewcommand{\infRule}[2]{#1 \,\Rightarrow\, #2}
\FIGtpisyntaxsemantics

Fig.~\ref{fig:pi/types+rules} presents the syntax
and semantics of the $\pi$-calculus.
Processes are built from \metaS names.
The process ``$\cIn{\cxc}{\cxx_1,\ldots,\cxx_k}.\cxP$'', which
(input)-binds the names $\cxx_i$'s, waits to receive a $k$-tuple of
names over channel $\cxc$ and then behaves like $\cxP$ with the received
values substituted for $\cxx_i$'s.
The process ``$\cOut{\cxc}{\cxx_1,\ldots,\cxx_k}.\cxP$'' sends the $k$-tuple
$\cxx_1$, $\ldots$, $\cxx_k$ over channel $\cxc$ and then behaves like $\cxP$.
Other constructors have the meaning as in \metaS (Sec.~\ref{sec:star/calculus}).
The sets of names $\fn{\cxP}$, $\fbn{\cxP}$, $\ibn{\cxP}$, $\nbn{\cxP}$
are defined as in \metaS.

Processes are identified up to $\alpha$-conversion of bound names
which preserves basic names.
A substitution in the $\pi$-calculus is a finite function from names to names,
and its application to $\cxP$ is written postfix, e.g.,
``$\sub[\{\cxx\mapsto\cxy\}]{\cxP}$''.
A process $\cxP$ is \defthis{well scoped} when \rulename{(S1)} 
$\fbn{\cxP}$, $\ibn{\cxP}$, and $\nbn{\cxP}$ do not overlap,
\rulename{(S2)} nested input binders do not bind the same basic name, and
\rulename{(S3)} no input action contains the same basic name more then once.
Henceforth, we require processes to be well scoped
(well-scopedness is preserved by rewriting).

\begin{example}\label{ex:pi/calculus}
Let 
    $\cxP = \oB\cIn{\n{s}}{\n{x},\n{y}}.\cOut{\n{x}}{\n{y}}.\pNull\oI
           \cOut{\n{s}}{\n{a},\n{n}}.\pNull  \hspace{0.4mm} \oI
           \cIn{\n{a}}{\n{v}}.\cIn{\n{v}}{\n{p}}.\pNull \hspace{3.6mm} \oI
           \cOut{\n{n}}{\n{o}}.\pNull                          \; \oI$ \\ 
    $ \hspace*{55.2mm}
      \oI \cOut{\n{s}}{\n{b},\n{m}}.\pNull
      \oI \cIn{\n{b}}{\n{w}}.\cIn{\n{v}}{\n{q},\n{r}}.\pNull
      \oI \cOut{\n{m}}{\n{o},\n{o}}.\pNull 
    $

\noindent
Using the rewriting relation $\SYMgRew$ sequentially four times we can
obtain (among others) the process
    ``\begin{math}
        \oB\cIn{\n{s}}{\n{x},\n{y}}.\cOut{\n{x}}{\n{y}}.\pNull  \oI
        \cIn{\n{n}}{\n{p}}.\pNull                               \oI
        \cOut{\n{n}}{\n{o}}.\pNull                              \oI
        \cIn{\n{m}}{\n{q},\n{r}}.\pNull                         \oI
        \cOut{\n{m}}{\n{o},\n{o}}.\pNull 
    \end{math}''.
\end{example}

\subsection{Types for the Polyadic $\pi$-calculus (\TPi)}
\label{sec:pi/system}

\renewcommand{\infAxiom}[1]{#1}
\renewcommand{\infRule}[2]{#1\,\Rightarrow\,#2}
\FIGtpitypesrules

We compare \polyS with a simple type system \cite[Ch. 3]{Tur:PhD-1995}
for the polyadic $\pi$-calculus presented by Turner which we name \TPi.
\TPi is essentially Milner's sort discipline \cite{Mil:CMS-1999}.
In the polyadic settings, an arity mismatch error on
channel $\cxc$ can occur when the lengths of the sent and received tuple do
not agree, like in
``$\cIn{\cxc}{\cxx}.\pNull\oI\cOut{\cxc}{\cxy,\cxy}.\pNull$''.
Processes which can never evolve to a state with a similar situation are
called \defthis{communication safe}.
\TPi verifies communication safety of $\pi$-processes.

The syntax and typing rules of \TPi are presented in
Fig.~\ref{fig:pi/types+rules}.
Recall that $\base{\cxx}$ denotes the basic name of $\cxx$.
Types $\cxT$ are assigned to names.
Type variables $\cxTV$ are types of names which are not used as channel
labels.
The type ``$\cChT{\cxT_1,\ldots,\cxT_k}$'' describes a channel which
can be used to communicate any $k$-tuple whose $i$-th name has type $\cxT_i$.
A context $\cxE$ assigns types to free names of a process (via their basic
names).
%\footnote{Turner's original system does not use \metaS basic
%names and assigns types directly to names. This technical variation simplifies 
%the correspondence with \polyS.}). 
The relation $\cType{\cxP}$, which is preserved under rewriting, expresses that
the actual usage of channels in $\cxP$ agrees with $\cxE$.
When $\cType{\cxP}$ for some $\cxE$ then $\cxP$ is communication safe.
The opposite does not necessarily hold.

\begin{example}
\label{ex:pi/system}

Given $\cxP$ from Ex.~\ref{ex:pi/calculus} we can see that there is
no $\cxE$ such that $\cType{\cxP}$. 
It is because the parts $\cOut{\n{s}}{\n{a},\n{n}}$ and
$\cOut{\n{s}}{\n{a},\n{m}}$ imply that types of $\n{n}$ and $\n{m}$ must be
equal while the parts $\cOut{\n{n}}{\n{o}}$ and $\cOut{\n{m}}{\n{o},\n{o}}$
force them to be different.
On the other hand $\cxP$ is communication safe.
We check this using \polyS in Sec~\ref{sec:pi/instatiate}.

\end{example}

\subsection{Instantiation of \metaS to the $\pi$-calculus}
\label{sec:pi/instatiate}

\newcommand{\FIGpiencoding}{
\begin{figure}[t]
    \small
    \vspace{-1mm}
    \renewcommand{\arraystretch}{1.2}
    \setlength{\arraycolsep}{1mm}
    \begin{displaymath}
    \begin{array}{ll}
        \encode{\pNull}        = \pNull
    & 
        \encode{\txP_0\oI \txP_1}  = \encode{\txP_0} \oI
            \encode{\txP_1}
    \\
        \encode{\oB\txP} = \oB\encode{\txP}
    & 
        \encode{\cIn{\cxc}{\cxx_1,\ldots,\cxx_k}.\cxP} =
            \sChIn{\cxc}{\cxx_1,\ldots,\cxx_k}.\encode{\cxP}
    \\
        \encode{\cNu{\cxx}{\cxP}} = \pNu{\cxx}{\encode{\txP}}\
    & 
        \encode{\cOut{\cxc}{\cxx_1,\ldots,\cxx_k}.\cxP} =
            \sChOut{\cxc}{\cxx_1,\ldots,\cxx_k}.\encode{\cxP}
    \end{array}
    \end{displaymath}
    \vspace{-1mm}
    \renewcommand{\arraystretch}{1}
    \normalsize
    \caption{Encoding of $\pi$-calculus processes in \metaS.}
    \label{fig:pi/process+encoding}
\end{figure}
}

\newcommand{\EXpiprincipaltype}{
    \newlength{\col}
    \newlength{\row}
    \setlength{\col}{5mm}
    \setlength{\row}{-4pt}
    \newcommand*{\armAa}{1.00}
    \newcommand*{\armAb}{1.45}
    \newcommand*{\armAc}{0.40}
    \newcommand*{\armAd}{0.92}
    \newcommand*{\armAe}{0.40}
    \newcommand*{\armAf}{0.90}
    \small
    \begin{displaymath}
    \begin{psmatrix}[colsep=\col,rowsep=\row]
    %1         2         3          4         5         6         7         8          9
            & \tick   &          &  &          &  &         &          &         \\ % 1
            &         &          &  &          &  &         &          &         \\ % 2
            & \tick   &          &  &          &  &         & \tick    &         \\ % 3
            &         &          &  &          &  &         &          &         \\ % 4
    \tick   &         & \tick    &  &          &  &         & \tick    &         \\ % 5
            &         &          &  &          &  &         &          &         \\ % 6 
            &         &          &  & \node{R} &  & \tick   &          & \tick   \\ % 7 
            &         &          &  &          &  &         &          &         \\ % 8 
    \tick   &         & \tick    &  &          &  &         & \tick    &         \\ % 9 
            &         &          &  &          &  &         &          &         \\ % 10
            & \tick   &          &  &          &  &         & \tick    &         \\ % 11
            &         &          &  &          &  &         &          &         \\ % 12
            & \tick   &          &  &          &  &         &          &         \\ % 13
        \psset{arrows=->,nodesep=2pt,labelsep=1pt}
        \everypsbox{\scriptstyle}
        % s(x,y).x<y>
        \ncline{7,5}{7,7}\naput[nrot=:U]{\sChIn{\n{s}}{\n{x},\n{y}}}
        \ncline{7,7}{7,9}\naput[nrot=:U]{\sChOut{\n{x}}{\n{y}}}
        % a(v).v(p)
        \ncline{7,5}{5,3}\nbput[nrot=:D]{\sChIn{\n{a}}{\n{v}}}
        \ncline{5,3}{5,1}\nbput[nrot=:D]{\sChIn{\n{v}}{\n{p}}}
        % b(w).w(q,r)
        \ncline{7,5}{9,3}\nbput[nrot=:D]{\sChIn{\n{b}}{\n{w}}}
        \ncline{9,3}{9,1}\nbput[nrot=:D,npos=0.4]{\sChIn{\n{w}}{\n{q},\n{r}}}
        % n(p)
        \ncdiagg[angleA=125,armA=\armAa,linearc=.3]{7,5}{3,2}
            \nbput[nrot=:D,npos=1.45]{\sChIn{\n{n}}{\n{p}}}
        % n<o>
        \ncdiagg[angleA=115,armA=\armAb,linearc=.3]{7,5}{1,2}
            \nbput[nrot=:D,npos=1.45]{\sChOut{\n{n}}{\n{o}}}
        % m(q,r)
        \ncdiagg[angleA=235,armA=\armAa,linearc=.3]{7,5}{11,2}
            \nbput[nrot=:D,npos=1.45]{\sChIn{\n{m}}{\n{q},\n{r}}}
        % m<o,o>
        \ncdiagg[angleA=245,armA=\armAb,linearc=.3]{7,5}{13,2}
            \nbput[nrot=:D,npos=1.45]{\sChOut{\n{m}}{\n{o},\n{o}}}
        % s<a,n>
        \ncdiagg[angleA=55,armA=\armAc,linearc=.3]{7,5}{5,8}
            \naput[nrot=:U,npos=1.5]{\sChOut{\n{s}}{\n{a},\n{n}}}
        % a<n>
        \ncdiagg[angleA=65,armA=\armAd,linearc=.3]{7,5}{3,8}
            \naput[nrot=:U,npos=1.5]{\sChOut{\n{a}}{\n{n}}}
        % s<b,m>
        \ncdiagg[angleA=-55,armA=\armAe,linearc=.3]{7,5}{9,8}
            \naput[nrot=:U,npos=1.5]{\sChOut{\n{s}}{\n{b},\n{m}}}
        % b<m>
        \ncdiagg[angleA=-65,armA=\armAf,linearc=.3]{7,5}{11,8}
            \naput[nrot=:U,npos=1.5]{\sChOut{\n{b}}{\n{m}}}
    \end{psmatrix}
    \end{displaymath}
    \normalsize
}

The $\pi$-calculus syntax from Sec.~\ref{sec:pi/calculus} already
matches the \metaS syntax and thus only the following $\xRSP$ is needed
to instantiate \metaS to the calculus $C_\xRSP$ and \polyS to its type
system $S_\xRSP$.
Sec.~\ref{sec:pi/embedding} shows that $C_\xRSP$ is essentially
identical to the above $\pi$-calculus.
%
%\vspace{-10pt}
\PARinstpi
Each communication prefix length has its own rule; in our
implementation, a single rule can uniformly handle all lengths, but
the formal \metaS presentation is deliberately simpler.
The next example shows how to check communication safety in $S_\xRSP$
without using \TPi.

\begin{wraptable}{r}{5cm}
\vspace{-2mm}
\centering
\EXpiprincipaltype
\vspace{-13mm}
\end{wraptable}

\begin{fakedexample}
\label{ex:pi/instatiate}
    Let $\xP$ be a \metaS equivalent of $\cxP$ from Ex.~\ref{ex:pi/calculus}.
    We can compute a principal $\xRSP$-type $\xS_\xP$ of $\xP$ which is
    displayed on the right.
    Node $\node{R}$ is its root. 
    The type $\xS_\xP$ contains all computational futures of $\xP$ in one
    place.
    Thus, because there are no two edges from $\node{R}$ labeled by
    ``$\sChIn{\xa}{\xab_1,\ldots,\xab_k}$'' and
    ``$\sChOut{\xa}{\xab_1',\ldots,\xab_j'}$'' with $k\ne j$, we can conclude
    that $\xP$ is communication safe which Ex.~\ref{ex:pi/system} shows \TPi
    can not do.
    Our implementation can be instructed (using an additional rule) to
    insert the error name $\sBullet$ at the place of communication errors.
    Any type of $\xP$ without $\sBullet$ then implies $\xP$'s communication
    safety.

\end{fakedexample}

\subsection{Embedding of \TPi in \polyS}
\label{sec:pi/embedding}

\newcommand{\FIGpitypeembed}{
\begin{figure}[t]
\small
\labeledHeader{The set of expected and actual channel types of $\xG$:}
\vspace{1mm}
\begin{displaymath}
\begin{array}{l}
    \chtypes{\cxE}{\!\xG} \!=\!
    \{
        ( \cxE(\xa), \cChT{\cxE(\xab_1),\ldots,\cxE(\xab_k)} ) \!\such\!
        (\sEdge{\xNodeX\!}{\!\sChIn{a}{\xab_1,\ldots,\xab_k\!}}{\!\xNodeY})
        \!\in\!\xG
        \!\lor\!
        (\sEdge{\xNodeX\!}{\!\sChOut{a}{\xab_1,\ldots,\xab_k\!}}{\!\xNodeY})
        \!\in\!\xG
    \} 
\end{array}
\end{displaymath}
\vspace{-2mm}
\labeledHline{Context $\cxE$ and shape type $\xS$ agreement relation
$\SYMcAgree$:}
\vspace{1mm}

Write $\cAgree{\cxE}{\sShape{\xG}{\xNode}}$ when there is some $\cxE'$ with
the domain disjoint from $\cxE$ such that $\chtypes{\cxE\cup\cxE'}{\xG}$ is
defined and is an identity.

\normalsize
\caption{Property of shape types corresponding to $\vdash$ of \TPi.}
\label{fig:pi/type+embedding} 
\end{figure}
}

Using the terminology from Sec.~\ref{sec:general/terminology} we have
that $C$ is the $\pi$-calculus, $S_C$ is \TPi,
predicates $\gxT$ of $S_C$ are contexts $\cxE$, and $S_C$'s
relation $\gType{\gxP}{\gxT}$ is $\cType{\cxP}$.
Moreover $\xRS$ is $\xRSP$ which was introduced with 
$C_\xRSP$ and $S_\xRSP$ in Sec.~\ref{sec:pi/instatiate}.
This section provides a formal comparison which shows how to, for a
given $\cxP$ and $\cxE$, answer the question $\cType{\cxP}$ using
$S_\xRSP$.

As stated in Sec.~\ref{sec:general/calculus}, to relate \TPi and
$S_\xRSP$ we need to provide an encoding $\encode{\cdot}$ of
$\pi$-processes in \metaS.
This $\encode{\cdot}$, found in TR
    \cite[Fig.~10]{Jak+Wel:ShapeTypes-2009}
, is almost an identity because the
$\pi$-calculus syntax (Fig.~\ref{fig:pi/syntax+semantics}) already agrees
with \metaS.
Thus $\encode{\cdot}$ mainly changes the syntactic category.
Prop.~\ref{thm:general/encoding+correct} holds in the above context.

%\FIGpiencoding
\FIGpitypeembed

Given $\cxE$, we define a shape type property which
holds for the principal type $\xS_\cxP$ of $\encode{\cxP}$ iff
$\cType{\cxP}$.
The property is given by the relation $\cAgree{\cxE}{\xS}$ from
Fig.~\ref{fig:pi/type+embedding}.
The set $\chtypes{\cxE}{\xG}$ contains pairs of \TPi types
extracted from $\xG$.
Each pair corresponds to an edge of $\xG$ labeled by an
action type ``$\sChIn{\xa}{\xab_1,\ldots,\xab_k}$'' or
``$\sChOut{\xa}{\xab_1,\ldots,\xab_k}$''.
The first member of the pair is $\xa$'s type expected by $\cxE$,
and the second member computes $\xa$'s actual usage from the types of
$\xab_i$'s.
The set $\chtypes{\cxE}{\xG}$ is undefined when some required value of
$\cxE$ is not defined.
The context $\cxE'$ from the definition of $\SYMcAgree$ provides types
of names originally bound in $\bxP$.
These are not mentioned by $\cxE$ but are in $\xG$.
The following theorem shows how to answer $\cType{\cxP}$ by
$\SYMcAgree$.

\begin{theorem}\label{thm:pi/embedding+correct}
    
Let no two different binders in $\cxP$ bind the same basic name, $\xS_\cxP$
be a principal \mbox{($\xRSP$-)type} of $\encode{\cxP}$, and
$\dom{\cxE}=\fbn{\cxP}$.  
Then $\cType{\cxP}$ iff $\cAgree{\cxE}{\xS_\cxP}$.

\end{theorem}

The requirement on different binders (which can be achieved by renaming) is
not preserved under rewriting because replication can introduce two
same-named binders.
However, when all binding basic names differ in $\cxP_0$, then the
theorem holds for any successor $\cxP_1$ of $\cxP_0$ even when the
requirement is not met for $\cxP_1$.
We want to ensure that the derivation of $\cType{\cxP}$ does not assign
different types to different bound names.
A slightly stronger assumption of Thm.~\ref{thm:pi/embedding+correct}
simplifies its formulation.
The theorem uses principal types and does not necessarily hold for a
non-principal $\xRSP$-type $\xS$ of $\encode{\cxP}$ because $\xS$'s
additional edges not needed to match $\encode{\cxP}$ can preclude
$\cAgree{\cxE}{\xS}$.

\subsection{Conclusions}
\label{sec:pi/conclusion}

We showed a process (Ex.~\ref{ex:pi/calculus}) that can not be proved
communication safe by \TPi (Ex.~\ref{ex:pi/system}) but can be proved so
by \polyS (Ex.~\ref{ex:pi/instatiate}).
Thm.~\ref{thm:pi/embedding+correct} implies that \polyS recognizes safety
of all \TPi-safe processes.
Thus we conclude that \polyS is better in recognition of communication
safety then \TPi.
Thm.~\ref{thm:pi/embedding+correct} allows to recognize typability in
\TPi: $\cxP$ is typable in \TPi iff $\cAgree{\emptyset}{\xS_\cxP}$.
This is computable because a \polyS principal type can always be found (for
$S_\xRSP$ in
polynomial time), and checking $\SYMcAgree$ is easy.

Turner \cite[Ch.~5]{Tur:PhD-1995} presents also a polymorphic system for
the $\pi$-calculus which recognizes $\cxP$ from Ex.~\ref{ex:pi/calculus}
as safe.
However, with respect to our best knowledge, it can not recognize safety
of the process ``$\cxP\oI\cOut{\n{s}}{\n{n},\n{a}}.\pNull$'' which
\polyS can do.
We are not aware of any process that can be recognized safe by Turner's
polymorphic system but not by \polyS.
It must be noted, there are still processes which \polyS can not prove
safe, for example,
``$\cIn{\n{a}}{\n{x}}.\cIn{\n{a}}{\n{y},\n{z}}.\pNull \oI
   \cOut{\n{a}}{\n{o}}.\cOut{\n{a}}{\n{o},\n{o}}.\pNull$''.

Other $\pi$-calculus type systems are found in the literature.
Kobayashi and Igarashi \cite{Iga+Kob:POPL-2001} present types for the
$\pi$-calculus looking like simplified processes which can verify
properties which are hard to express using shape types (race conditions,
deadlock detection) but do not support polymorphism.
One can expect applications where \polyS is more expressive as well as
contrariwise.
Shape types, however, work for many process calculi, not just
the $\pi$-calculus.

\section{Shape Types for Mobile Ambients}
\label{sec:ma}

\subsection{Mobile Ambients (\MA)}
\label{sec:ma/calculus}

\newcommand{\FIGtmasyntaxsemantics}{
\begin{figure}[t]
\small
\labeledHeader{Syntax of \MA processes:}
\begin{pstgrammar}
    \grmeqclass{\txx}{AName}{\grmset{Name}\setminus
        \{ \sBullet \}} \\
    \grmclass{\txM}{ACapability}{
        \tNullMsg \grmOr
        \txx \grmOr
        \cab{in}{\txM} \grmOr
        \cab{out}{\txM} \grmOr
        \cab{open}{\txM} \grmOr
        \txM.\txM'
    } \\
    \grmclass{\txW}{AMessageType }{ 
        \mbox{definition postponed to Fig.~\ref{fig:ma/types+rules}}
    } \\
    \grmclass{\txP}{AProcess}{
        \pNull \grmOr
        (\txPP \oI \txPQ) \grmOr
        \amb{\txM}{\txP} \grmOr
        \txM.\txP \grmOr
        \oB \txP \grmOr
        \tNu{\txx}{\txW}{\txP} \grmOr
    } \\
    \grmcont{
        \tOut \grmOr
        \tIn{\txP}
    } 
\end{pstgrammar}
\setlength{\arraycolsep}{1mm}
\labeledHline{Rewriting relation of \MA
    ($\SYMbStreq$ is standard defined in TR
    \cite[Fig.~12]{Jak+Wel:ShapeTypes-2009}):}
\begin{displaymath}
\begin{array}{rcl}
    \amb{\txx}{\cab{in}{\txy}.\txPP\oI\txPQ}\oI\amb{\txy}{\txPR}  
    & \SYMtRewrite &
    \amb{\txy}{\amb{\txx}{\txPP\oI\txPQ}\oI\txPR} 
\\
    \amb{\txy}{\amb{\txx}{\cab{out}{\txy}.\txPP\oI\txPQ}\oI\txPR}  
    & \SYMtRewrite &
    \amb{\txx}{\txPP\oI\txPQ}\oI\amb{\txy}{\txPR} 
\\
    \cab{open}{\txx}.\txPP\oI\amb{\txx}{\txPQ}  
    & \SYMtRewrite &
    \txPP\oI\txPQ 
\\
    \tIn{\txP}\oI\tOut
    & \SYMtRewrite &
    \tSub[\{\txx_1\mapsto\txM_1,\ldots,\txx_k\mapsto\txM_k\}]{\txP}
\end{array}
\end{displaymath}
\begin{displaymath}
\begin{array}{ll}
    \infRule{ \tRewrite{\txPP}{\txPQ} }
            { \tRewrite{\amb{\txx}{\txPP}}{\amb{\txx}{\txPQ}} }
  &
    \infRule{ \tRewrite{\txPP}{\txPQ} }
            { \tRewrite{\tNu{\txx}{\txW}{\txPP}}
                       {\tNu{\txx}{\txW}{\txPQ}} }
\\
    \infRule{ \tRewrite{\txPP}{\txPQ} }
            { \tRewrite{\txPP\oI\txPR}{\txPQ\oI\txPR} }
    \quad
  &
    \infRule{ \tStreq{\txPP'}{\txPP} \land
                 \tRewrite{\txPP}{\txPQ} \land
                 \tStreq{\txPQ}{\txPQ'} }
            { \tRewrite{\txPP'}{\txPQ'} }
\end{array}
\end{displaymath}
\vspace{-2mm}

\normalsize
\caption{Syntax and semantics of \TMA.}
\label{fig:ma/syntax+semantics}
\end{figure}
}

\newcommand{\FIGtmatypesrules}{
\begin{figure}[t]
\small
\labeledHeader{Syntax of \TMA types:}
\begin{pstgrammar}
    \grmclass{\txW}{AMessageType }{ \tAmb{\txT} \grmOr \tCap{\txT} } \\
    \grmclass{\txT}{AExchangeType}{ \tShh \grmOr \tExch } \\
    \grmeqclass{\txE}{AEnvironment}{ 
        \mapfin{\grmset{AName}}{\grmset{AMessageType}} } \\
\end{pstgrammar}
\labeledHline{Typing rules of \TMA:}
\begin{displaymath}
\begin{array}{ll}
    \begin{array}{l}
        \infRule{ \tEnv{\txx} =  \txW }
                { \tType{\txx}{\txW}  }
    \\
        \infRule{ \tType{\txM}{\tAmb{\txTS}} }
                { \tType{\cab{in}{\txM}}{\tCap{\txTT}} }
    \\
        \infRule{ \tType{\txM}{\tAmb{\txTS}} }
                { \tType{\cab{out}{\txM}}{\tCap{\txTT}} }
    \\
        \infRule{ \tType{\txM}{\tAmb{\txT}} }
                { \tType{\cab{open}{\txM}}{\tCap{\txT}} }
    \end{array}
  &
    \begin{array}{l}
        \infAxiom{ \tType{\tNullMsg}{\tCap{\txT}} }
    \\
        \infRule{ \tType{\txM}{\tCap{\txT}} \land
                  \tType{\txM'}{\tCap{\txT}} }
                { \quad \\ \hfill \tType{\txM.\txM'}{\tCap{\txT}} }
    \end{array}
\end{array}
\end{displaymath}
\begin{displaymath}
\setlength{\arraycolsep}{2mm}
\begin{array}{ll}
    \begin{array}{l}
        \infRule{ \tType{\txP}{\txT} }
                { \tType{\oB\txP}{\txT} }
    \\
        \infAxiom{ \tType{\pNull}{\txT} }
    \end{array}
  &
    \begin{array}{l}
        \infRule{ \tType{\txM}{\tCap{\txT}} \land
                  \tType{\txP}{\txT} }
                { \tType{\txM.\txP}{\txT} }
    \\
        \infRule{ \tType{\txM}{\tAmb{\txT}} \land
                  \tType{\txP}{\txTT} }
                { \tType{\amb{\txM}{\txP}}{\txTS} }
    \\
        \infRule{ \tType{\txPP}{\txT} \land \tType{\txPQ}{\txT} }
                { \tType{\txPP\oI\txPQ}{\txT} }
    \end{array}
\end{array}
\end{displaymath}
\begin{displaymath}
\begin{array}{l}
        \infRule{ \tType[\extend{\txE}
                        {\txx\mapsto\tAmb{\txTS}}]{\txP}{\txTT} }
                { \tType{\tNu{\txx}{\tAmb{\txTS}}{\txP}}{\txTT} }
    \\
        \infRule{ \forall i\!: 0<i\le k \land \tType{\txM_i}{\txW_i} }
                { \tType{\tOut}{\tExch} }
    \\
        \infRule{ \tType[\extend{\txE}{\txx_1\mapsto\txW_1,\ldots,
                                       \txx_k\mapsto\txW_k}]
                        {\txP}{\tExch} }
                { \\ \qquad \tType{\tIn{\txP}}{\tExch} }
\end{array}
\end{displaymath}
\vspace{-2mm}
\normalsize
\caption{Syntax of \TMA types and typing rules.}
\label{fig:ma/types+rules}
\end{figure}
}

\newcommand{\DEFtmawellscoped}{
\begin{definition}
Call $\txP$ or $\txE$  \defthis{well formed} when all of the following hold:
\begin{description}
\item[\rulename{(S1)}] $\fbn{\txP}$, $\ibbn{\txP}$, and $\nbbn{\txP}$ are
    pairwise disjoint
\item[\rulename{(S2)}] for 
    $\sIn{\isa{\txa_1^{i_1}}{\txW_1},\ldots,
          \isa{\txa_k^{i_k}}{\txW_k}}.\txP_0$
    in $\txP$,
    $\txa_j$'s are distinct and %not bound in $\txP_0$
        $\txa_j\not\in\ibbn{\txP_0}$
\item[\rulename{(S3)}] different binding occurrences of ``$\txa$'' assign the
    same type to ``$\txa$''
\item[\rulename{(S4)}] $\txE$ assigns the same type to names which share a
    basic name
\end{description}
\end{definition}
}

Mobile Ambients (\MA), introduced by Cardelli and Gordon
\cite{Car+Gor:FoSSaCS-1998}, is a process calculus for representing process
mobility.
Processes are placed inside named bounded locations called \emph{ambients}
which form a tree hierarchy.
Processes can change the hierarchy and send messages to nearby processes.
Messages contain either ambient names or hierarchy change instructions.

\renewcommand{\infAxiom}[1]{#1}
\renewcommand{\infCondAxiom}[2]{#2 & \mbox{if} & #1}
\renewcommand{\infRule}[2]{#1 \,\Rightarrow\, #2}
\FIGtmasyntaxsemantics

Fig.~\ref{fig:ma/syntax+semantics} describes \MA process syntax.
%Some names, e.g., ``$\n{in}$'', are reserved for the translation.
Executing a capability consumes it and instructs the surrounding ambient
to change the hierarchy.
The capability ``$\cab{in}{\txx}$'' causes moving into a sibling ambient
named $\txx$, the capability ``$\cab{out}{\txx}$'' causes moving out of
the parent ambient $\txx$ and becoming its sibling, and
``$\cab{open}{\txx}$'' causes dissolving the boundary of a child ambient
$\txx$.
In capability sequences, the left-most capability will be executed first.

The constructors ``$\pNull$'', ``$\oI$'', ``$.$'', ``$\oB$'', and
``$\nu$'' have standard meanings.
Binders contain explicit type annotations (Sec.~\ref{sec:ma/system} below).
The expression $\amb{\txx}{\txP}$ describes the process $\txP$ running
inside ambient $\txx$.
Capabilities can be communicated in messages.
$\sOut{\txM_1,\ldots,\txM_k}$ is a process that sends a $k$-tuple of
messages.
$\tIn{\txP}$ is a process that receives a $k$-tuple of messages, substitutes
them for appropriate $\txx_i$'s in $\txP$, and continues as this new process.
Free and bound (basic) names are defined like in \metaS.
Processes that are $\alpha$-convertible are identified.
A substitution $\xSub$ is a finite function from names to messages and
its application to $\txP$ is written $\sub{\txP}$.
Fig.~\ref{fig:ma/syntax+semantics} also describes structural equivalence and
semantics of \MA processes.
The only thing the semantics does with type annotations is copy them around.
We require all processes to be well-scoped w.r.t.\ conditions
\rulename{S1-3} from Sec.~\ref{sec:pi/calculus}, and the additional
condition \rulename{(S4)} that the same message type is assigned to
bound names with the same basic name.
Ambients and capabilities where $\txM$ is not a single name, which the
presentation allows for simplicity, are inert and meaningless.

\begin{example}
\label{ex:ma/calculus}
In this example, packet ambient $\n{p}$ delivers a
synchronization message to destination ambient $\n{d}$ by following
instructions $\n{x}$. 
As we have not yet properly defined message types, we only suppose
$\txW_\n{p}=\tAmb{\txT}$ for some $\txT$.

\vspace{-4mm}
\small
\begin{displaymath}
\begin{array}{ll}
    \txP = &
        \sOut{\cab{in}{\n{d}}} \oI
        \tNu{\n{p}}{\txW_\n{p}}{(
            \amb{\n{d}}{\cab{open}{\n{p}}.\pNull} \oI
            \sIn{\isa{\n{x}}{\txW_\n{x}}}.\amb{\n{p}}{\n{x}.\sOut{}} 
        )} \SYMtRewrite
\\ &
        \tNu{\n{p}}{\txW_\n{p}}{(
            \amb{\n{d}}{\cab{open}{\n{p}}.\pNull} \oI
            \amb{\n{p}}{\cab{in}{\n{d}}.\sOut{}} 
        )} \SYMtRewrite
%\\ &
        \tNu{\n{p}}{\txW_\n{p}}{(
            \amb{\n{d}}{\cab{open}{\n{p}}.\pNull\oI\amb{\n{p}}{\sOut{}}}
        )} \SYMtRewrite
            \amb{\n{d}}{\sOut{}}
\end{array}
\end{displaymath}
\normalsize
\end{example}

\subsection{Types for Mobile Ambients (\TMA)}
\label{sec:ma/system}

An arity mismatch error, like in
   ``$\sOut{\n{a},\n{b}}.\pNull \oI \sIn{\n{x}}.\cab{in}{\n{x}}.\pNull$'',
can occur in polyadic \MA.
Another communication error can be encountered when a sender sends
a capability while a receiver expects a single name.
For example
    ``$\sOut{\cab{in}{\n{a}}}.\pNull \oI \sIn{\n{x}}.\cab{out}{\n{x}}.\pNull$''
can rewrite to a meaningless ``$\cab{out}{(\cab{in}{\n{a}})}.\pNull$''.
Yet another error happens when a process is to
execute a single name capability, like in ``$\n{a}.\pNull$''.
Processes which can never evolve to a state with any of the above
errors are called \emph{communication safe}.
A typed \MA introduced by Cardelli and Gordon
\cite{Car+Gor:POPL-1999}, which we name \TMA, verifies communication safety.

\renewcommand{\infAxiom}[1]{#1}
\renewcommand{\infRule}[2]{#1\mathbin{\Rightarrow}#2}
\FIGtmatypesrules

\TMA assigns an allowed communication topic to each ambient location and
ensures that processes respect the topics.
Fig.~\ref{fig:ma/types+rules} describes \TMA type syntax.
Exchange types, which describe communication topics, are assigned to
processes and ambient locations.
The type $\tShh$ indicates silence (no communication).
$\tExch$ indicates communication of $k$-tuples of
messages whose $i$-th member has the message type $\txW_i$.
For $k=0$ we write $\tOne$ which allows only
synchronization actions $\sOut{}$ and $\sIn{}$.
$\tAmb{\txT}$ is the type of an ambient where communication
described by $\txT$ is allowed.
$\tCap{\txT}$ describes capabilities whose execution can unleash
exchange $\txT$ (by opening some ambient).
Environments assign message types to free names (via basic names).
Fig.~\ref{fig:ma/types+rules} also describes the \TMA
typing rules.
Types from conclusions not mentioned in the assumption can be arbitrary.
For example, the type of $\amb{\txM}{\txP}$ can be arbitrary provided
$\txP$ is well-typed.
It reflects the fact that the communication inside $\txM$ does not
directly interact with $\txM$'s outside.
Existence of some $\txE$ and $\txT$ such that $\txE$ does not assign a
$\textsf{Cap}$-type to any free name and $\tType{\txP}{\txT}$ holds implies
that $\txP$ is communication safe.

\begin{example}
\label{ex:ma/system}
Take $\txP$ from Ex.~\ref{ex:ma/calculus},
    $\txE = \{ 
              \n{d}\mapsto\tAmb{\tOne}
            \}$, and
    $\txW_\n{p} = \tAmb{\tOne}$, and
    $\txW_\n{x} = \tCap{\tOne}$.
We can see that $\tType{\txP}{\tCap{\tOne}}$ but, for example,
$\tType[\txE\not]{\txP}\tOne$.  
\end{example}

\subsection{Instantiation of \metaS to \MA}
\label{sec:ma/instatiate}

\newcommand{\FIGtmaencoding}{
\begin{figure}[t]
\small
\vspace{-3mm}
\begin{displaymath}
\begin{array}{lll}
    \cabname{\txM} = \begin{cases}
        \txx     & \mbox{if } \txM=\txx\in\grmset{AName}\quad \\
        \sBullet & \mbox{otherwise}
    \end{cases}
&
    \begin{array}{ll}
        \cabenc{\cab{in}{\txM}}   = \cab{in}{\cabname{\txM}}     \\
        \cabenc{\cab{out}{\txM}}  = \cab{out}{\cabname{\txM}}    \\
        \cabenc{\cab{open}{\txM}} = \cab{open}{\cabname{\txM}}\quad  
    \end{array}
&
    \begin{array}{ll}
        \cabenc{\tNullMsg}     = \sNull                    \\
        \cabenc{\txM_0.\txM_1} = \cabenc{\txM_0}.\cabenc{\txM_1}   
    \end{array}
\end{array}
\end{displaymath}
\vspace{-3mm}
\renewcommand{\arraystretch}{1.2}
\begin{displaymath}
\begin{array}{ll}
    \encode{\pNull}        = \pNull
&
    \encode{\txP_0\oI \txP_1}  = \encode{\txP_0} \oI
        \encode{\txP_1}
\\
    \encode{\oB\txP} = \oB\encode{\txP}
&
    \encode{\sOut{\txM_1,\ldots,\txM_k}} =
        \sOut{\cabenc{\txM_1},\ldots,\cabenc{\txM_k}}.\pNull
\\
    \encode{\txM.\txP} = \splice{\cabenc{\txM}}{\encode{\txP}}
&
    \encode{(\nu\isa{\x}{W})\txP} = \pNu{\x}{\encode{\txP}}\
\\
    \encode{\amb{\txM}{\txP}} = \amb{\cabname{\txM}}{\encode{\txP}}\qquad
&
    \encode{\sIn{\isa{\x_1}{W_1},\ldots,\isa{\x_k}{W_k}}.\txP} = 
       \sIn{\x_1,\ldots,\x_k}.\encode{\txP}
\end{array}
\end{displaymath}
\vspace{-2mm}
\renewcommand{\arraystretch}{1}
\normalsize
\caption{Encoding of \TMA processes in \metaS.}
\label{fig:ma/processencoding}
\end{figure}
}

\newcommand{\PARinstma}{
\small
\begin{displaymath}
    \begin{array}{rl}
        \xRSA = \big\{ 
        & \sActive{\var{P}}{\amb{\var{a}}{\var{P}}},
    \\
        & \sReduce{ 
            \amb{\var{a}}{\n{in}\ \var{b}.\var{P}\oI\var{Q}}
            \oI \amb{\var{b}}{\var{R}}
          }{
            \amb{\var{b}}{
                \amb{\var{a}}{\var{P}\oI\var{Q}} 
                \oI \var{R}}
          },
    \\
        & \sReduce{
            \amb{\var{a}}{\amb{
                \var{b}}{\n{out}\ \var{a}.\var{P}\oI\var{Q}}\oI\var{R}}
          }{
            \amb{\var{a}}{\var{R}}
            \oI \amb{\var{b}}{\var{P}\oI\var{Q}}
          },
    \\
        & \sReduce{
            \n{open}\ \var{a}.\var{P}\oI\amb{\var{a}}{\var{R}}
          }{
            \var{P}\oI\var{R}
          }
          \;\big\}\;\cup
    \\
        \bigcup_{k=0}^{\infty} \big\{ & %\; & 
        \oReduce\oLeft\,
            \sOut{\var{M}_1,\ldots,\var{M}_k}.\var{P} \oI
            \sIn{\var{a}_1,\ldots,\var{a}_k}.\var{Q} 
            \oArr\;
            \var{P}\oI\oLeft
                \var{a}_1\!\oAssign\var{M}_1,\ldots,
                \var{a}_k\!\oAssign\var{M}_k
            \oRight\,\var{Q}
            \,\oRight
        \;\big\}
    \end{array}
\end{displaymath}
\normalsize
}

\newcommand{\EXtmainstatiate}{
    \setlength{\col}{8mm}
    \setlength{\row}{8mm}
    \small
    \begin{displaymath}
    \begin{psmatrix}[colsep=\col,rowsep=\row]
    % 1     2       3          4       5     
    \tick &       & \node{R} & \tick & \tick \\ % 1
          & \tick &          & \tick & \tick \\ % 2
    \tick & \tick & \tick    &       & \tick    % 3
        \psset{arrows=->,nodesep=2pt,labelsep=1pt}
        \everypsbox{\scriptstyle}
        % d[]
        \ncline{1,3}{2,2}\nbput[nrot=:D]{\amb{\n{d}}{}}
        % inside of `d[]` 
        \ncline{2,2}{3,3}\naput[nrot=:U]{\amb{\n{p}}{}}
        \ncline{2,2}{2,4}\naput[nrot=:U]{\cab{in}{\n{d}}}
        \ncline{2,2}{3,1}\nbput[nrot=:D]{\cab{open}{\n{p}}}
        \ncline{2,2}{3,2}\nbput[nrot=:D]{\sOut{}}
        % p[].<>
        \ncline{1,3}{2,4}\naput[nrot=:U]{\amb{\n{p}}{}}
        \ncline{2,4}{3,5}\naput[nrot=:U]{\sOut{}}
        % inside of `d[].p[]`
        \ncline{3,3}{2,4}\naput[nrot=:U]{\cab{in}{\n{d}}}
        \ncline{3,3}{3,5}\naput[nrot=:U]{\sOut{}}
        % <{in d}*>
        \ncline{1,3}{1,1}\nbput[nrot=:D]{\sOut{\sStar{\{\cab{in}{\n{d}}\}}}}
        %(x).p[].x.<>
        \ncline{1,3}{1,4}\naput[nrot=:U]{\sIn{\n{x}}}
        \ncline{1,4}{1,5}\naput[nrot=:U]{\amb{\n{p}}{}}
        \ncline{1,5}{2,5}\naput[nrot=:U]{\n{x}}
        \ncline{2,5}{3,5}\naput[nrot=:U]{\sOut{}}
        % `d[]` and `in d` loops
        \nccurve[angleA=90,angleB=180,ncurv=11]{2,2}{2,2}
            \nbput[nrot=:D]{\amb{\n{d}}{}}
        \nccurve[angleA=90,angleB=0,ncurv=11]{2,4}{2,4}
            \naput[nrot=:U]{\cab{in}{\n{d}}}
    \end{psmatrix}
    \end{displaymath}
    \normalsize
}

When we omit type annotations, add ``$\pNull$'' after output actions, and write capability prefixes always in a right
associative manner (like
    ``$\cab{in}{\n{a}}.(\cab{out}{\n{b}}.(\cab{in}{\n{c}}.\pNull))$''), 
we see that the \MA syntax is included in the \metaS syntax.
The following set $\xRSA$ instantiates \metaS to \MA.
% 
% \vspace{-2mm}
\PARinstma
%\vspace{-1mm}
The $\oActive$ rule lets rewriting be done inside ambients.
It corresponds to the rule
    ``$\tRewrite{\txPP}{\txPQ}\Rightarrow
      \tRewrite{\amb{\txx}{\txPP}}{\amb{\txx}{\txPQ}}$''.
Each communication prefix length has its own rule as in the case of the
$\pi$-calculus.
$\xRSA$ defines the calculus $C_\xRSA$ and the type system $S_\xRSA$.

Communication safety of $\xP$ can be checked on an $\xRSA$-type as follows.
Two edges with the same source labeled by $\sIn{\xa_1,\ldots,\xa_k}$ and
$\sOut{\xab_1,\ldots,\xab_j}$ with $k\ne j$
indicates an arity mismatch error (but only at active positions).
Every label containing $\sBullet$ (introduced by a
substitution) indicates that a capability was sent instead of a name.
Moreover, an edge labeled with a name $\xa\not\in\ibn{\xP}$ at active position
indicates an execution of a single name capability.
A type of $\xP$ not indicating any error proves $\xP$'s safety.
Checking safety this way is easy.

\begin{wraptable}{r}{4cm}
    \vspace{-5mm}
    \centering
    \EXtmainstatiate
    \vspace{-15mm}
\end{wraptable}

\begin{fakedexample}
\label{ex:ma/instatiate}
    $C_\xRSA$'s equivalent of $\txP$ from Ex.~\ref{ex:ma/calculus} is
    $\xP =
        \sOut{\cab{in}{\n{d}}}.\pNull \oI
        \pNu{\n{p}}{(
            \amb{\n{d}}{\cab{open}{\n{p}}.\pNull} \oI
            \sIn{\n{x}}.\amb{\n{p}}{\n{x}.\sOut{}.\pNull} 
        )}$.
    Its principal $\xRSA$-type is displayed on the right.
    Its root is $\node{R}$ and other node names are omitted. 
    Checking the edge labels as described above easily proves $\xP$'s
    safety.
    The edge labeled by $\n{x}$ is not a communication error because $\n{x}$
    is input-bound in $\xP$.

\end{fakedexample}

\subsection{Embedding of \TMA in \polyS}
\label{sec:ma/embedding}

\newcommand{\FIGtmaembedding}{
\begin{figure}[t]
\small
\labeledHeader{Extraction of types of bound names:}
    \vspace{1mm}
    \begin{displaymath}
    \begin{array}{lcl}
            \tInEnv(\xa) = \txW & \mbox{iff} & 
            \txP\mbox{ has a subprocess }
                \sIn{\ldots,\isa{\xa^i}{\txW},\ldots}.\txP_0 
        \\[1mm]
            \tNuEnv(\xa)  =  \txW & \mbox{iff} & 
            \txW\!=\!\tAmb{\txT} \!\land\!
            \txP\mbox{ has a subprocess }\tNu{\xa^i}{\txW}{\txP_0} 
    \end{array}
    \end{displaymath}
    \vspace{.5mm}

\labeledHline{Type information:}
    \vspace{1mm}
    \begin{pstgrammar}
        \grmeqclass{\xI}{TypeInfo}{
            \grmset{AEnvironment}\times
            \grmset{AEnvironment}\times
            \grmset{AExchangeType}    
        }
    \end{pstgrammar}
    For a given $\xI = (\txE_0,\txE_1,\txT)$ we write $\SYMtiAll$ for
    $\txE_0\cup\txE_1$, and $\SYMtiCom$ for $\txE_1$, and $\tiType$ for $\txT$.
    \vspace{.5mm}
 
\labeledHline{Set of nodes of a shape graph (and correspondence functions):}
    \begin{displaymath}
        \extypes = \{ \tiType \} \cup
        \{ \txT \such \tAmb{\txT} \in \rng{\SYMtiAll} \}
    \qquad
        \SYMnodeof = \invert{\SYMtypeof}
    \end{displaymath}
    Let $\nodes$ be an arbitrary but fixed set of nodes such that there exist
    the bijection $\SYMtypeof$ from $\nodes$ into $\extypes$.
    \vspace{1mm}

\labeledHline{Action types describing legal capabilities:}
\vspace{.5mm}
\begin{displaymath}
\begin{array}{l}
    \namesof{\txW} = \{ \xa \such \tiAll{\xa}=\txW \}
%\\
    \qquad
    \allowedin{\txT} = \moves \cup \opens{\txT} \cup \comm{\txT}
\\[1mm]
    \moves = \{ \cab{in}{\xa}, \cab{out}{\xa} \such
        \exists\txT.\,\xa\in\namesof{\tAmb{\txT}} \}
\\
    \opens{\txT} = \{ \cab{open}{\xa} \such 
            \xa\in\namesof{\tAmb{\txT}} \} \cup
    \namesof{\tCap{\txT}}
\\
    \msg{\tAmb{\txT}} = \namesof{\tAmb{\txT}}
\\
    \msg{\tCap{\txT}} = \namesof{\tCap{\txT}} \cup
        \{ \sStar{(\moves\cup\opens{\txT})} \}
\\
    \comm{\tShh} = \emptyset 
\qquad
    \comm{\tExch} = \{ \sOut{\xMT_1,\ldots,\xMT_k} \such
                       \xMT_i\in\msg{\txW_i} \} \cup
\\ \hspace{35mm}
                    \{ \sIn{\xa_1,\ldots,\xa_k} \such
                       \tiCom{\xa_i} = \txW_i \land
                       (i\ne j\Rightarrow \xa_i\ne\xa_j)                           
                       \}
\end{array}
\end{displaymath}
%\vspace{1mm}
\labeledHline{Construction of shape predicates:}
\begin{displaymath}
\begin{array}{ll}
    \typenc{\xI} = 
    \sShape{\typegraph{\xI}}{\nodeof{\tiType}}
    \hspace{2mm}
&
    \typegraph{\xI} = 
    \{ \sEdge{\xNode}{\ \xAT\ }{\xNode} \!\such\!
       \xAT\!\in\!\allowedin{\typeof{\xNode}} \!\land\!
       \xNode\!\in\!\nodes
    \} \ \cup
\\ & \hspace{2mm}
    \{ \sEdge{\xNodeX}{\amb{\xa}{}}{\xNodeY} \!\such\!
       \xa\!\in\!\namesof{\tAmb{\typeof{\xNodeY}}} \!\land\!
       \xNodeX,\xNodeY\!\in\!\nodes
    \} 
\end{array}
\end{displaymath}
\vspace{-1mm}
\normalsize
\caption{Construction of \polyS type embedding.}
\label{fig:ma/embedding}
\end{figure}
}

Using the notation from Sec.\ref{sec:general/terminology} we have that
$C$ is \MA, $S_C$ is \TMA, predicates $\gxT$ are pairs $(\txE,\txT)$, and 
$S_C$'s relation $\gType{\gxP}{\gxT}$ is $\tType{\txP}{\txT}$.
Moreover $\xRS$ is $\xRSA$ which was introduced with $C_\xRSA$ and $S_\xRSA$
in Sec.~\ref{sec:ma/instatiate}.
This section provides an embedding which
shows how to, for a given $\txP$, $\txE$, and $\txT$, answer the question
$\tType{\txP}{\txT}$ using $S_\xRSA$.
We stress that it is primarily a theoretical embedding for proving greater
expressiveness which is not intended for use in practice.

%\FIGtmaencoding

%Fig.~\ref{fig:ma/processencoding} provides an encoding $\encode{\cdot}$
%of \MA processes in \metaS.
%The encoding $\encode{\cdot}$ translates capabilities to \metaS messages
%and \MA processes to \metaS processes.

An encoding $\encode{\cdot}$ of \MA processes in \metaS, found in TR
    \cite[Fig.~14]{Jak+Wel:ShapeTypes-2009}, is
again almost an identity except for the following.
(1) Meaningless expressions allowed by \MA's syntax are translated using 
the special name $\sBullet$, e.g.,
    ``$\cabenc{\cab{in}{(\cab{out}{\n{a}})}}=\cab{in}{\sBullet}$''.
(2) The encoding erases type annotations which is okay because \MA's
rewriting rules only copy them around.
The type embedding below recovers type information by different means.
Prop.~\ref{thm:general/encoding+correct} holds in the given context.

As discussed in Sec.~\ref{sec:general/system}, we can not translate
$(\txE,\txT)$ to a shape type with an equivalent meaning because
$\vdash$ is preserved under renaming of bound basic names.
%because the
%relation $\vdash$ is preserved under renaming of bound basic names.
Nevertheless this becomes possible when we specify the sets of allowed
input- and $\nu$-bound basic names and their types.
These can be easily extracted from a given process $\txP$.
An environment $\tNuEnv$ (resp. $\tInEnv$) from the top part of
Fig.~\ref{fig:ma/embedding} describes $\nu$-bound (resp. input-bound)
basic names of $\txP$.
The definition reflects that $\nu$-bound names in typable processes can
only have $\textsf{Amb}$-types.
For a given $\txE$, $\txP$, and $\txT$ we construct the shape type
$\typenc{\txE\cup\tNuEnv,\tInEnv,\txT}$ such that $\tType{\txP}{\txT}$ 
iff
$\shapes{\encode{\txP}}{\typenc{\txE\cup\tNuEnv,\tInEnv,\txT}}$.
The construction needs to know which names are input-bound and thus they
are separated from the other names.
The well-scopedness rules S1-4 ensure that there is no ambiguity in
using only basic names to refer to typed names in a process.
The type information $\xI$ (Fig.~\ref{fig:ma/embedding}, 2nd part)
collects what is needed to construct a shape type.
For $\xI=(\txE\cup\tNuEnv,\tInEnv,\txT)$ we define $\SYMtiAll$, $\SYMtiCom$,
and $\tiType$ such that 
$\SYMtiAll$ describes types of all names in $\txE$ and $\txP$, and
$\SYMtiCom$ describes types of $\txP$'s input-bound names, and $\tiType$
is simply $\txT$.

\begin{example}
    $\txE$, $\txP$, and $\txT$ from the previous examples (Ex.~\ref{ex:ma/calculus} and
    Ex.~\ref{ex:ma/system}) give us  
    $\xI = (\txE\cup\tNuEnv,\tInEnv,\tCap{\tOne})$ and we have:
    \vspace{-2mm}
    \small
    \begin{displaymath}
        \txE\cup\tNuEnv = \{
            \n{d}\mapsto\tAmb{\tOne},
            \n{p}\mapsto\tAmb{\tOne}
        \}
        \quad
        \SYMtiCom = \{
            \n{x}\mapsto{\tCap{\tOne}}
        \}
        \quad
        \SYMtiAll = \txE\cup\tNuEnv\cup\SYMtiCom
    \end{displaymath}
    \normalsize
\end{example}

The main idea of the construction of the shape type $\typenc{\xI}$ from
$\xI$ is that $\typenc{\xI}$ contains exactly one node for every
exchange type of some ambient location, that is, one node for the
top-level type $\tiType$, and one node for $\txTS$ whenever $\tAmb{\txTS}$
is in $\xI$.
The top-level type corresponds to the shape type root.
Each node corresponding to some $\txT$ has self-loops which describe
all capabilities and communication actions which a process of the type
$\txT$ can execute.
When $\tiAll{\n{d}}=\tAmb{\tOne}$ then every node would have a self-loop
labeled by ``$\cab{in}{\n{d}}$'' because $\n{in}$-capabilities can be
executed by any process.
On the other hand only the node corresponding to $\tOne$ would allow
``$\cab{open}{\n{d}}$'' because only processes of type $\tOne$ can legally
execute it.
Finally, following an edge labeled with ``$\amb{\n{d}}{}$'' means entering $\n{d}$.
Thus the edge has led to the node $\xNode_\n{d}$ that corresponds to $\tOne$.
In the above example, the shape graph would contain edges labeled with 
``$\amb{\n{d}}{}$'' from any node to $\xNode_\n{d}$.

\FIGtmaembedding

The construction starts by building the node set of a shape predicate
(Fig.~\ref{fig:ma/embedding}, 3rd part).
All the exchange types of ambient locations are gathered in the set
$\extypes$.
These types are put in bijective correspondence with the set $\nodes$.

\begin{example}
Our example gives 
$\extypes = 
    \{ \tCap{\tOne}, \tOne \}$.
Let us take $\nodes=\{\node{R},\node{1}\}$ and define the bijections
such that 
    $\nodeof{\tCap{\tOne}}=\node{R}$ and
    $\nodeof{\tOne}=\node{1}$.
\end{example}

The 4th part of Fig.~\ref{fig:ma/embedding} defines some auxiliary
functions.
The set $\namesof{\txW}$ contains all basic names declared with the type
$\txW$ by $\xI$.  
The set $\allowedin{\txT}$ contains all \polyS action types which
describe (translations of) all capabilities and action prefixes which
are allowed to be legally executed by a process of the type $\txT$.
The set $\allowedin{\txT}$ consists of three parts: $\moves$,
$\opens{\txT}$, and $\comm{\txT}$.
The action types in $\moves$ describe all $\n{in}/\n{out}$ capabilities
constructible from ambient basic names in $\xI$.
The set does not depend on $\txT$ because $\n{in}/\n{out}$ capabilities can
be executed by any process.
The set $\opens{\txT}$ describe $\n{open}$-capabilities 
which can be executed by a process of the type $\txT$.
The second part of $\opens{\txT}$ describes names of the type
$\tCap{\txT}$ which might be instantiated to some executable capabilities.
The set $\comm{\txT}$ describes communication actions which can be executed
by a process of the type $\txT$.
Its first part describes output- and the second input-actions.
The auxiliary set $\msg{\txW}$ describes all messages of the type $\txW$
constructible from names in $\xI$.

\begin{example}Relevant sets for our example are:
%
%\vspace{-2mm}
\small
\begin{displaymath}
\begin{array}{ll}
    \namesof{\tAmb{\tOne}} = \{ \n{d}, \n{p} \} \qquad
  &
    \opens{\tOne} = \{ \cab{open}{\n{d}}, \cab{open}{\n{p}}, \n{x} \}
\\
    \namesof{\tCap{\tOne}} = \{ \n{x} \}
  &
    \opens{\tCap{\tOne}} = \emptyset   
\\
    \comm{\tOne} = \{ \sOut{}, \sIn{} \}
&
    \moves = \{ 
        \cab{in}{\n{d}}, \cab{in}{\n{p}},
        \cab{out}{\n{d}}, \cab{out}{\n{p}}
    \}
\\
  \multicolumn{2}{l}{ 
    \comm{\tCap{\tOne}} = \{
        \sOut{\n{x}},
        \sOut{\sStar{\{ 
            \cab{in}{\n{d}}, \cab{in}{\n{p}},
            \cab{out}{\n{d}}, \cab{out}{\n{p}},
            \cab{open}{\n{d}}, \cab{open}{\n{p}}, 
            \n{x}         
        \}}},
        \sIn{\n{x}}
    \}
  }
\end{array}
\end{displaymath}
\normalsize
\end{example}

The bottom part of Fig.~\ref{fig:ma/embedding} constructs the shape
graph $\typegraph{\xI}$ and the shape predicate $\typenc{\xI}$ from
$\xI$.
The first part of $\typegraph{\xI}$ describes self-loops of $\xNode$ which
describe actions allowed to be executed by a process of
$\typeof{\xNode}$.
The second part of $\typegraph{\xI}$ describe transitions among nodes.
Any edge labeled by ``$\amb{\xa}{}$'' always leads to
the node which corresponds to the exchange type allowed inside $\xa$.

\begin{example}
\newcommand*{\RNodeLabel}{
    \begin{array}{l}
        \scriptstyle
        \cab{in}{\n{d}}  \oI
        \cab{out}{\n{d}} \oI
        \cab{in}{\n{p}}  \oI
        \cab{out}{\n{p}} \oI
        \sOut{\n{x}}     \oI
        \sIn{\n{x}}      \oI
    \\
        \scriptstyle
        \sOut{\sStar{\{
            \n{x},
            \cab{in}{\n{d}}, 
            \cab{in}{\n{p}}, 
            \cab{out}{\n{p}}, 
    \\
        \scriptstyle\qquad
            \cab{out}{\n{d}}, 
            \cab{open}{\n{d}}, 
            \cab{open}{\n{p}}
        \}}}
    \end{array}
}
\newcommand*{\ONodeLabel}{
    \begin{array}{llllll}
        \scriptstyle \cab{in}{\n{d}}   & \scriptstyle \mathop{\oI} &
        \scriptstyle \cab{out}{\n{d}}  & \scriptstyle \mathop{\oI} &
        \scriptstyle \cab{open}{\n{d}} & \scriptstyle \mathop{\oI}
    \\
        \scriptstyle \cab{in}{\n{p}}   & \scriptstyle \mathop{\oI} &
        \scriptstyle \cab{out}{\n{p}}  & \scriptstyle \mathop{\oI} &
        \scriptstyle \cab{open}{\n{p}} & \scriptstyle \mathop{\oI}
    \\
        \multicolumn{5}{l}{
           \scriptstyle 
           \n{x}          \hfill\mathop{\oI}\hfill 
           \sOut{}        \hfill\mathop{\oI}\hfill 
           \sIn{}         \hfill\mathop{\oI}\hfill 
           \amb{\n{p}}{}  \hfill\mathop{\oI}\hfill 
           \amb{\n{d}}{}
        }
    \end{array}
}

The resulting shape predicate $\typenc{\xI}=\sShape{\xG}{\node{R}}$ in our
example is as follows.
We merge edges with the same source and destination using ``$\oI$''.

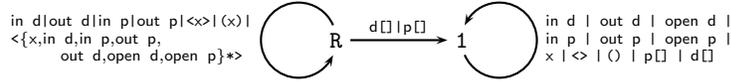
\small
\renewcommand{\arraystretch}{0.6}
\begin{displaymath}
\begin{psmatrix}[colsep=1.5cm]
        \node{R} & \node{1}
    \psset{arrows=->,nodesep=3pt,labelsep=1pt}
    \everypsbox{\scriptstyle}
    \ncline{1,1}{1,2}\naput{\amb{\n{d}}{}\oI\amb{\n{p}}{}}
    \rput{L}{\nccircle{1,1}{.5cm}\nbput[nrot=R]{\RNodeLabel}}
    \rput{R}{\nccircle{1,2}{.5cm}\nbput[nrot=L]{\ONodeLabel}}
\end{psmatrix}
\end{displaymath}
\normalsize
\renewcommand{\arraystretch}{1.0}

\end{example}

Correctness of the translation is expressed by
Thm.~\ref{thm:ma/embedding+correct}.
The assumptions ensure that no $\nu$-bound name is mentioned by
$\txE$ or has a $\textsf{Cap}$-type assigned by an annotation.
Here we just claim that $\typenc{\xI}$ is always an $\xRSA$-type.

\begin{theorem}\label{thm:ma/embedding+correct}
    Let $\dom{\txE}\cap\nbn{\txP}=\emptyset$ and $\dom{\tNuEnv}=\nbn{\txP}$.
    Then it holds that
    $\tType{\txP}{\txT}$ if and only if 
    $\shapes{\encode{\txP}}{
        \typenc{(\txE\cup\tNuEnv,\tInEnv,\txT)}}$.
\end{theorem}

\subsection{Conclusions}
\label{sec:ma/conclusion}

We embedded \TMA's typing relation in $S_\xRSA$
(Sec.~\ref{sec:ma/embedding}) and showed how to recognize communication
safety in $S_\xRSA$ directly (Sec.~\ref{sec:ma/instatiate}).
The type $\typenc{\xI}$ constructed in Sec.~\ref{sec:ma/embedding} can
also be used to prove the safety of $\txP$.
But then, it follows from the properties of principal types, that the safety
of $\txP$ can be recognized directly from its principal $\xRSA$-type. 
Thus any process proved safe by \TMA can be proved safe by $S_\xRSA$ on
its own.
    
Some processes are recognized safe by $S_\xRSA$ but not by \TMA.
For example, 
    ``$\sIn{\isa{\n{x}\!}{\!\txW}}.\n{x}.\pNull \oI \sOut{\cab{in}{\n{a}}}$'' 
is not typable in \TMA but it is trivially safe.
Another examples show polymorphic abilities of shape types,
for example, the $C_\xRSA$ process
\vspace{-2mm}
\begin{displaymath}
    \oB\sIn{\n{x},\n{y},\n{m}}.\amb{\n{x}}{\cab{in}{\n{y}}.\sOut{\n{m}}.\pNull} \oI 
    \sOut{\n{p},\n{a},\n{c}}.\pNull \oI
    \amb{\n{a}}{\cab{open}{\n{p}}.\pNull} \oI
    \sOut{\n{q},\n{b},\cab{in}{\n{a}}}.\pNull \oI 
    \amb{\n{b}}{\cab{open}{\n{q}}.\pNull}
\end{displaymath}
can be proved safe by \polyS but it constitutes a challenge for \TMA-like
non-polymorphic type systems.
We are not aware of other type systems for \MA and its successors that can
handle this kind of polymorphism.

The expressiveness of shape types $\typenc{\xI}$ from
Sec.~\ref{sec:ma/embedding} can be improved.
In subsequent work \cite{Car+Ghe+Gor:ICALP-1999}, Cardelli, Ghelli, and
Gordon define a type system which can ensure that some ambients stay
immobile or that their boundaries are never dissolved.
This can be achieved easily by removing appropriate self loops of nodes.
We can also assign nodes to (groups of) ambients instead of exchange
types. 
This gives us similar possibilities as another \TMA successor
\cite{Car+Ghe+Gor:ICTCS-2000}.
Moreover, we can use shape type polymorphism to express
location-dependent properties of ambients, like that ambient $\n{a}$ can be
opened only inside ambient $\n{b}$.

\section{Conclusions and Future Work}
\label{sec:conclusion}

We discussed already the contributions (Sec.~\ref{sec:contributions},
\ref{sec:star/discussion}).
Conclusions for the embeddings were given separately
(Sec.~\ref{sec:pi/conclusion}, \ref{sec:ma/conclusion}).
Future work is as follows.
For extensions, priorities are better handling of choice (e.g., because of
its use in biological system modeling), and handling of
\textbf{rec} which is in many calculi more expressive
than replication and better describes recursive behavior.
Moreover we would like to generalize actions so that calculi
with structured messages, like the Spi calculus
    \cite{Aba+Gor:IAC-1999}, can be handled.
For applications, we would like to (1) relate shape types with 
other systems which also use graphs to represent types
    \cite{Yoshida:FSTTCS-1996,Konig:CONCUR-1999},
and (2) to study the relationship between shape types and session types
\cite{Hon:CONCUR-1993}.

\bibliography{polystar-comparison}

\begin{thebibliography}{10}

\bibitem{Car+Ghe+Gor:ICALP-1999}
L.~Cardelli, G.~Ghelli, and A.~D. Gordon.
\newblock Mobility types for mobile ambients.
\newblock In {\em ICALP}, volume 1644 of {\em LNCS}, pages 230--239, July 1999.

\bibitem{Car+Ghe+Gor:ICTCS-2000}
L.~Cardelli, G.~Ghelli, and A.~D. Gordon.
\newblock Ambient groups and mobility types.
\newblock In {\em IFIP TCS}, volume 1872 of {\em LNCS}, pages 333--347, Aug.
  2000.

\bibitem{Car+Gor:FoSSaCS-1998}
L.~Cardelli and A.~D. Gordon.
\newblock Mobile ambients.
\newblock In {\em FoSSaCS}, volume 1378 of {\em LNCS}, pages 140--155, 1998.

\bibitem{Car+Gor:POPL-1999}
L.~Cardelli and A.~D. Gordon.
\newblock Types for mobile ambients.
\newblock In {\em POPL}, pages 79--92, 1999.

\bibitem{Aba+Gor:IAC-1999}
M.~A.~D. Gordon.
\newblock A calculus for cryptographic protocols: The spi calculus.
\newblock {\em Inf.\ {\&} Comp.}, 148(1):1--70, Jan. 1999.

\bibitem{Hon:CONCUR-1993}
K.~Honda.
\newblock Types for dyadic interaction.
\newblock In {\em CONCUR}, volume 715 of {\em LNCS}, pages 509--523, 1993.

\bibitem{Iga+Kob:POPL-2001}
A.~Igarashi and N.~Kobayashi.
\newblock A generic type system for the pi-calculus.
\newblock In {\em POPL}, pages 128--141, 2001.

\bibitem{Jak:SYR-2009}
J.~Jakub\r{u}v.
\newblock {\em A Second Year Report}.
\newblock Heriot-Watt Univ., MACS., 2009.
\newblock Available at http://www.macs.hw.ac.uk/{\textasciitilde}jj36.

\bibitem{Jak+Wel:ShapeTypes-2009}
J.~Jakub\r{u}v and J.~B. Wells.
\newblock The expressiveness of generic process shape types.
\newblock Technical Report HW-MACS-TR-0069, Heriot-Watt Univ., July 2009.

\bibitem{Konig:CONCUR-1999}
B.~K{\"o}nig.
\newblock Generating type systems for process graphs.
\newblock In {\em CONCUR}, volume 1664 of {\em LNCS}, pages 352--367, 1999.

\bibitem{Mak+Wel:PolyStar-2004}
H.~Makholm and J.~B. Wells.
\newblock Instant polymorphic type systems for mobile process calculi: Just add
  reduction rules and close.
\newblock Technical Report HW-MACS-TR-0022, Heriot-Watt Univ., Nov. 2004.

\bibitem{Mak+Wel:ESOP-2005}
H.~Makholm and J.~B. Wells.
\newblock Instant polymorphic type systems for mobile process calculi: Just add
  reduction rules and close.
\newblock In {\em ESOP}, volume 3444 of {\em LNCS}, pages 389--407, 2005.

\bibitem{Mil:CMS-1999}
R.~Milner.
\newblock {\em Communicating and Mobile Systems: The $\pi$-Calculus}.
\newblock Cambridge Press, 1999.

\bibitem{Mil+Par+Wal:IC-1992}
R.~Milner, J.~Parrow, and D.~Walker.
\newblock A calculus of mobile processes.
\newblock {\em Inf.\ {\&} Comp.}, 100(1):1--77, Sept. 1992.

\bibitem{Nie+Nie+Pri+Ros:ENTCS-2007-v180n3}
F.~Nielson, H.~R. Nielson, C.~Priami, and D.~Rosa.
\newblock Control flow analysis for bioambients.
\newblock {\em ENTCS}, 180(3):65--79, 2007.

\bibitem{Tur:PhD-1995}
D.~N. Turner.
\newblock {\em The Polymorphic Pi-Calculus: Theory and Implementation}.
\newblock PhD thesis, Uni.\ of Edinburgh, 1995.
\newblock Rep.\ ECS-LFCS-96-345.

\bibitem{Wells:ICALP-2002}
J.~B. Wells.
\newblock The essence of principal typings.
\newblock In {\em ICALP}, volume 2380 of {\em LNCS}, pages 913--925, 2002.

\bibitem{Yoshida:FSTTCS-1996}
N.~Yoshida.
\newblock Graph types for monadic mobile processes.
\newblock In {\em FSTTCS}, volume 1180 of {\em LNCS}, pages 371--386, 1996.

\end{thebibliography}

\end{document}